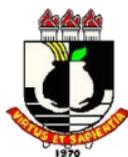

UNIVERSIDADE FEDERAL DE MATO GROSSO
INSTITUTO DE BIOCIÊNCIAS
CURSO DE MESTRADO EM ECOLOGIA E CONSERVAÇÃO DA
BIODIVERSIDADE

**DISTRIBUIÇÃO E DIVERSIDADE DE HERBÁCEAS DE SUB-BOSQUE EM UMA
FLORESTA DE TERRA FIRME DA AMAZÔNIA MERIDIONAL**

**ELIANA CELESTINO DA PAIXÃO RODRIGUES DOS SANTOS**

CUIABÁ-MT
2012

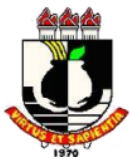

UNIVERSIDADE FEDERAL DE MATO GROSSO
INSTITUTO DE BIOCIÊNCIAS
CURSO DE MESTRADO EM ECOLOGIA E CONSERVAÇÃO DA
BIODIVERSIDADE

**DISTRIBUIÇÃO E DIVERSIDADE DE HERBÁCEAS DE SUB-BOSQUE EM UMA FLORESTA DE TERRA FIRME DA AMAZÔNIA MERIDIONAL**

**ELIANA CELESTINO DA PAIXÃO RODRIGUES DOS SANTOS**

> Dissertação apresentada ao Curso de Pós-Graduação, do Instituto de Biociências, como requisito parcial para obtenção do título de mestre em Ecologia e Conservação da Biodiversidade.

CUIABÁ-MT
2012

**Orientadora:** Profª. Dra. CÁTIA NUNES DA CUNHA
UFMT – Universidade Federal de Mato Grosso
IB – Instituto de Biociências

**Co-orientadora:** Profª. Dra. FLÁVIA R. C. COSTA
Programa de Pesquisa em Biodiversidade
INPA – Instituto Nacional de Pesquisas da Amazônia

**BANCA EXAMINADORA**

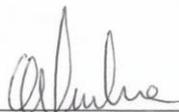

Profa. Dra. Cátia Nunes da Cunha

Universidade Federal de Mato Grosso

Orientadora

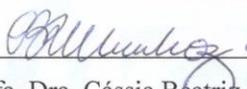

Profa. Dra. Cássia Beatriz Rodriguez Munhoz

Universidade de Brasília

Examinador Externo

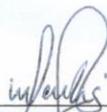

Prof. Dr. Jerry Magno Ferreira Penha

Universidade Federal de Mato Grosso

Examinador Titular

Prof. Dr. Domingos de Jesus Rodrigues

Universidade Federal de Mato Grosso

Examinador Suplente

## AGRADECIMENTOS

A CAPES pela bolsa de estudos, ao Programa de Pós-Graduação em Ecologia e Conservação da Biodiversidade da Universidade Federal de Mato Grosso.

Agradeço principalmente aos professores do programa, em especial, a minha orientadora profa. Doutora Cátia Nunes que esta me acompanhando desde os meus tempos remotos da graduação. Agradeço muito pela confiança, pelo apoio e incentivo em momentos oportunos, espero ter correspondido a todas suas expectativas, e continuar trabalhando ao seu lado.

A minha co-orientadora Profa. Doutora Flávia Costa pela ajuda nas análises, pelo treinamento para trabalhar com as ervas. Obrigada por tudo, inclusive por ter me feito amar esse mundo fascinante das ervas, antes por mim negligenciadas.

Ao professor Doutor Alexandre Salino, por ter me ajudando tanto com as samambaias, que deixou seus alunos de lado para sentar comigo e me ensinar cada detalhe importante na identificação das mesmas. Jamais esquecerei a atenção e a dedicação com que fui tratada por ele.

Ao professor Ms. Francisco Athayde pela ajuda nas identificações do material.

A Deus que me deu forças e sabedoria para fazer minhas escolhas, me protegendo sempre, em todos os momentos, principalmente nos campos!

Aos professores que aceitaram participar da minha defesa: Dra. Cássia, Dr. Jerry e Dr. Domingos, sei que poderão contribuir muito para o enriquecimento de meu trabalho.

A minha eterna amiga e companheira de campo Enildes (Nildes), sempre me transmitindo segurança e coragem mesmo diante dos brejos e jararacas. Parceira de todos os momentos, principalmente quando me atolava, estando ao meu lado e sempre rindo muito, é claro! Agradeço imensamente a ajuda no meu trabalho e tenho certeza que minhas coletas não seriam as mesmas sem sua presença.

Ao eterno amigo da eco-campo, Almir – Garfield – por me ajudar com os artigos, toda vez que eu precisava de artigos clássicos era a ele que eu atormentava e com muita reclamação e alegria ele sempre me ajudou. Obrigada de coração Garfield!

Aos amigos eternos de Manaus: Cadu e Gabriel, que incomodei bastante, Marcelo Xelento, por me abrigar em sua casa quando precisei e ao Paul por doar seu quarto pra mim e me ajudar com as compras, serei eternamente grata.

Ao Nandinho, eterno amigo a quem tenho uma enorme admiração e respeito, principalmente por ter perdido seu óculos num igarapé ao tentar coletar uma ervinha para mim! Minhas parcelas ripárias não seria as mesmas sem você.

Ao Ricardo Xexelento que me aturou, me ajudou, me aconselhou em momentos oportunos tanto em campo, feliz da vida por estar me ajudando nas parcelas ripárias, perdendo as lentes dos óculos pra me ajudar, e também no laboratório, é duro ter um amigo assim!

Ao pessoal do Laboratório de Ecologia Vegetal de Cuiabá: Cândida, eterna irmã e amiga, Thayse, Regiane, Abílio, Luciana e Sejana, meu muito obrigado, vocês sim me acolheram como uma família quando aí cheguei.

Ao pessoal do ABAM, que me apoiou quando precisei: Juliane, Márcia, Prof. Rafael, sempre atencioso e me dando forças e ânimos nos momentos difíceis, Prof. Domingos e Prof. Marlinton. Ao Everton pela confecção dos mapas, ao Robson por ter voltado até a parcela nove para pegar *Cyathea* fértil pra mim e pelas fotos, que precisavam ser perfeitas! Você é um show de emoções peão. Ao Lucas que me ajudou muito abrindo as trilhas para facilitar o trabalho. Ah! Se fosse o Lucas... já teria terminado!

A Rino Kids - Janinha que conseguiu morar comigo por quase um ano, Vitoriosa! Acordando cedo para me fazer companhia. Minha filha branca e mais velha...te gosto de mais.

A Patynha, a quem sempre serei grata pelos sábios conselhos, que me ajudaram muito em tomadas de decisões importantíssimas na minha vida.

Ao Rodrigo, a quem nem tenho palavras para agradecer tudo que fez por mim, mesmo não me suportando no início do curso, agora somos irmãos, te gosto muitão.

Ao meu Marido Nego – Reinaldo, que sempre me apoiou em todas as minhas escolhas, se mostrando um companheiro admiravelmente compreensivo, mesmo quando eu precisava viajar e ficar meses fora de casa!

As minhas filhas, Alexandra e Arielly que aceitaram ficar na casa de meus pais por dois anos até a conclusão do curso, e que agora não querem mais voltar pra casa, amo vocês, voltem pra casa!

Aos meus pais, quero que saibam que lamento muito por todos os momentos que não pude estar presente nesses dois anos, mas sei que vocês sempre torceram muito por mim, cuidado de minhas pérolas enquanto eu estava estudando.

A toda a galera da minha turma do curso de pós-graduação, que foram um show de emoções, principalmente aqueles com corações peludos e línguas pretas.

Agradeço ainda as donas serpentes por terem ficado em suas casinhas comportadinhas, enquanto eu fazia minhas amostragens.

Muito obrigada

# SUMÁRIO

# LISTA DE FIGURAS

# LISTA DE TABELAS

# RESUMO

A heterogeneidade ambiental é um fator determinante da estrutura das comunidades biológicas. Assim, compreender a distribuição das espécies ao longo de gradientes ambientais fornece subsídios para conservação. O objetivo deste trabalho foi determinar o padrão de distribuição da comunidade herbácea em três áreas da Amazônia Meridional. A amostragem foi realizada em 3 módulos totalizando 39 parcelas permanentes de acordo com o protocolo de coletas do Programa de Pesquisas em Biodiversidade. Todos os indivíduos herbáceos terrestres e hemiepífitos acima de 5 cm de altura foram amostrados. Análises multivariadas foram utilizadas para resumir a composição de espécies e modelos de regressão múltipla foram utilizados para determinar se, variáveis ambientais e de perturbação causada pela exploração de madeira influenciam na composição da comunidade herbácea. Foram inventariados 7.965 indivíduos, representando 70 espécies. A distância do curso d'água foi o principal fator associado à distribuição de espécies. Interações entre as variáveis mostrou que abertura de dossel e teor de areia também influenciam a composição de espécies, mas não houve efeito do número de árvores cortadas. A riqueza de espécies aumentou em áreas onde a abertura de dossel foi maior e diminui à medida que se distanciou do curso d'água. A ocorrência de hábitats preferenciais para algumas espécies tem, além de um interesse ecológico, um significado prático para a conservação e manejo destas espécies. Atualmente, a área de preservação permannete dos córregos prevista pelo código Florestal vigente é de 30 m para rios de até 10 m de largura, mas este estudo evidencia que a faixa de proteção deveria ser extendida para no mínimo 100 m de largura.

Palavras-chave: Ervas; pteridófitas; plantas de sub-bosque.

**ABSTRACT**

Environmental heterogeneity is a determining factor of the structure of biological communities. Thus, understanding the distribution of species along environmental gradients provides assistance to conservation. The goal of this study was to determine the distribution pattern of the herbaceous community in three areas of the Southern Amazon. Sampling was conducted in three modules totaling 39 permanent plots according to the protocol of collection of the Program for Research in Biodiversity. All herbaceous and ground hemiepiphyte subjects above 5 cm were recorded. Multivariate analyses were used to summarize the species composition, multiple regression models were used to determine if environmental variables and disturbance caused by logging influenced the composition of the herbaceous community. We recorded 7.965 individuals representing 70 species. The distance of the water course was the main factor associated with the distribution of the species, interactions between variables showed that canopy openness and sand content also influence the species composition, and there was no effect in the number of trees cut. Species richness increased in areas where canopy cover was higher and it decreases as it becomes more distant from the watercourse. The occurrences of preferred habitats for some species have, in addition to an ecological interest, a practical significance for the conservation and management of these species. Currently, the area of preservation of streams provided by the Forest Code in effect is 30 m for rivers up to 10 m wide. However, this study shows that the range of protection should be extended to at least 100 m wide.

Key-words: Herbs; pteridophytes; understory plants.

# 1. INTRODUÇÃO

A compreensão da variação florística em florestas tropicais apresenta uma importante aplicação prática em relação à conservação de biodiversidade, ao uso de recursos naturais e à restauração de áreas degradadas (CARVALHO, 2006; DRUCKER, 2005; VORMISTO et al., 2000). Para que se possa caracterizar e compreender a dinâmica nestes ambientes em toda a sua complexidade, são necessários estudos relacionados à composição florística e à estrutura fitossociológica de todos os estratos, inclusive o herbáceo (ANDRADE, 1992).

O estrato herbáceo das florestas tropicais tem revelado uma enorme diversidade (ANDREATA et al., 1997; CARVALHO, 2006; COSTA, 2004; GENTRY; DODSON, 1987; PRANCE, 1989; SOUZA et. al 2009). As ervas de sub-bosque funcionam como filtro ecológico em comunidades vegetais, modificando os diferentes ambientes de germinação e estabelecimento de mudas nas florestas, influenciando assim, composição e estrutura do dossel (HARMS et al., 2004; GILLIAN, 2007).

Dentre os reguladores da diversidade vegetal, solo e topografia são considerados os mais importantes por influenciar a distribuição de espécies e o padrão de comunidades em escalas locais (VORMISTO et al., 2000; CARVALHO, 2006) e regionais (TUOMISTO; POULSEN, 1996; DUQUE et al., 2002; TUOMISTO et al., 2003). Os estudos realizados procuram explicar as diferentes composições florísticas e estruturais do estrato herbáceo em diferentes tipos florestais, levando em consideração as exigências específicas das espécies quanto à luminosidade (BERNACCI, 1992; POULSEN; PENDRY, 1995), à disponibilidade de água (GENTRY; EMMONS, 1987; POULSEN, 1996), aos tipos de solos (GENTRY; EMMONS, 1987; POULSEN; BALSLEV, 1991; BERNACCI, 1992; RUOKOLAINEN; VORMISTO, 2000; TUOMISTO; POULSEN, 2000; TUOMISTO et al., 2003; COSTA et al., 2005; TUOMISTO; RUOKOLAINEN, 2005; TUOMISTO, 2006) e à altitude (GENTRY, 1988; COSTA et al., 2005;).

A umidade desempenha um papel fundamental na distribuição das plantas, principalmente para algumas espécies herbáceas de sub-bosque (PABST; SPIES, 1998) que são fortemente dependentes de locais úmidos para sua reprodução, (LINDH et al., 2003). A disponibilidade d'água é um importante preditor da composição da comunidade herbácea, visto que muitas dessas espécies ocorrem exclusivamente em áreas ripárias, onde a disponibilidade de água é maior, algumas espécies provavelmente não conseguiriam se estabelecer em outro ambiente (DRUCKER et al., 2008, GÁLHIDY et al., 2006 ), podendo ser consideradas especialistas. Esse recurso merece ser avaliado com atenção no que diz respeito à distribuição e riqueza de espécies herbáceas.

Existe uma forte influência da abertura de dossel na entrada de luz e consequentemente na distribuição das plantas de sub-bosque das florestas (SONG et al., 2004; VAN PELT; FRANKLIN, 2000). Como apenas 2% da luz solar incidente no dossel chegam ao sub-bosque de florestas tropicais (HOGAN; MACHADO, 2002; SVENNING, 1999), as plantas de sub-bosque respondem de maneira diferente aos padrões de luz que chegam até elas (VAN PELT; FRANKLIN, 2000). A luminosidade tem efeitos profundos sobre a composição e a riqueza de espécies herbáceas proporcionando uma maior taxa de germinação das sementes e crescimento de ervas no sub-bosque (GALHIDY et al., 2006), podendo ser um fator limitante para algumas espécies.

Algumas herbáceas apresentam maior abundância em locais onde a entrada de luz é maior, ambiente este que lhes permite melhores condições para competição, crescimento e reprodução (GALHIDY et al., 2006; SMITH, 1987). A quantidade de energia luminosa local é determinada por variações na abertura do dossel. Segundo Zuquim (2006) na maioria das vezes, áreas mais abertas estão associadas a clareiras naturais, cursos d'água e alterações humanas. No entanto, pouco se sabe sobre a influência relativa da luz na comunidade de ervas (mas veja JONES et al., 2006, ZUQUIM et al., 2009).

A textura do solo apresenta uma forte relação com a distribuição de espécies vegetais, sendo responsável pela retenção da umidade e nutrientes no solo, taxa de infiltração de água e drenagem e aeração (SOLLINS 1998), que afetam fortemente as plantas, influenciando a composição e a riqueza de espécies. Em florestas tropicais da Amazônia Central, a textura do solo é um dos fatores mais importantes na diferenciação florística (POULSEN; BALSLEV, 1991; CARVALHO, 2006; COSTA et al., 2005; KINUPP; MAGNUSSON, 2005; QUESADA et al., 2009; VORMISTO et al., 2000; ZUQUIM, 2006).

Além de fatores ambientais naturais, atividades antrópicas têm merecido atenção pelos impactos causados a essas comunidades. O impacto causado pela exploração de madeira pode afetar a composição e a riqueza de espécies (COSTA et al., 2002). Clareiras abertas pela retirada da madeira proporcionam maior disponibilidade de luz fazendo com que algumas espécies sejam totalmente excluídas enquanto outras aproveitam o espaço e as condições favoráveis para invadir o local impactado (FRANKLIN; VAN PELT, 2004; FREDERICKSEN; MOSTACEDO, 1999; VAN PELT; FRANKLIN, 1999; VELDMAN et al., 2009). SMITH (1987) sugeriu que as ervas de sub-bosque de florestas tropicais dependem de abertura no dossel para crescer e se reproduzir. Usando os níveis mais elevados de luz em clareiras, essas plantas devem apresentar uma alta saturação de luz para a fotossíntese. Algumas ervas de sub-bosque apresentam essa característica nas florestas do Panamá (SMITH, 1987). Se o mesmo mecanismo pode ser aplicado a outras florestas tropicais, esperamos que a composição e a riqueza de ervas sejam influenciadas pela extração de madeira.

Os impactos da exploração na composição de espécies geralmente são mais visíveis nos primeiros anos após a perturbação e tendem a desaparecer com o tempo, pois o dossel de florestas tropicais se desenvolve gradualmente em resposta ao aumento nos níveis de luz, mitigando assim vários efeitos causados pela exploração de madeira (BROADBENT et al., 2008; COSTA; MAGNUSSON, 2003; FRANKLIN; VAN PELT, 2004), mas alguns estudos também mostraram efeitos de longo prazo sobre a composição de espécies (FREDERICKSEN; MOSTACEDO, 1999; MAGNUSSON et. al, 1999; COSTA et al., 2002). Poucos estudos até o momento avaliaram o impacto da exploração seletiva de madeira sobre ervas de sub-bosque na Amazônia (COSTA; MAGNUSSON, 2002, 2003; COSTA et al., 2002) e, portanto não é possível afirmar ainda que esta comunidade não é negativamente afetada por este tipo de atividade em qualquer região da Amazônia.

A distribuição das espécies herbáceas e sub-arbustivas, bem como o crescimento e desenvolvimento dos indivíduos estão relacionados a uma interação complexa dos fatores citados. Esta complexidade que envolve tanto fatores naturais quanto antrópicos é comumente encontrada em florestas tropicais, porem suas interações são pouco estudadas. As espécies respondem de diferentes formas à todas essas características, que sem dúvida influenciam a composição de espécies nas florestas tropicais. Por esse motivo, este estudo teve como objetivos: a) Descrever a composição e a estrutura da comunidade herbácea em uma área com diferentes históricos de manejo da Amazônia Meridional, no norte do estado de Mato Grosso; b) avaliar os efeitos conjuntos do ambiente (dado pelas características do solo, topografia e luz) e do impacto antrópico provocado pela exploração madeireira na estruturação da comunidade de herbáceas em áreas da Amazônia Meridional.

## 2. ÁREAS DE ESTUDO

O estudo foi realizado em três áreas, sendo dois na Fazenda Continental (módulo1: 11º 34' 54.0''S e 55º 17' 15.6''W e módulo2: 11º 35' 20.3''S e 55º 17'34.7''W) e um na Fazenda Iracema (módulo 3: 11º 39'09.4''S e 55º 04' 54.6''W), ambas pertencentes ao município de Claudia – MT (Fig.1). De acordo com o sistema de Köppen o clima da região é classificado como Am, quente e úmido, com chuvas do tipo monçônico, uma transição entre o clima equatorial super-úmido (Af) da Amazônia e o tropical úmido (Aw) do Planalto Central. A região é caracterizada por duas estações, uma chuvosa, que ocorre de setembro a abril e concentra 80% das precipitações ocorridas durante o ano, e outra seca, que varia de maio a agosto, apresentando, nesse período, meses consecutivos com precipitação abaixo de 60 mm (MEDEIROS, 2004). A precipitação pluviométrica média anual na região é de 2.200 mm (VOURLITIS et al., 2002).

As áreas estudadas são compostas por florestas com diferentes históricos de manejo (corte seletivo de madeira), sendo que o módulo "1" foi manejado há nove anos (2002), o módulo "2" há 16 (1995) e o terceiro, há 30 anos (1981). Os módulos "1" e "2" foram instalados em uma área de floresta contínua, fragmentada apenas pela presença de estradas de terra desativadas ou em uso, para acesso à sede da fazenda, transporte de madeira e de produtos agrícolas. Já o módulo "3" foi estabelecido em um trecho de floresta cercada por lavouras, havendo ligação com duas grandes áreas de vegetação nativa. Pelo manejo realizado, em todas as áreas há presença de pequenas clareiras, ocasionadas pela retirada e armazenamento de madeira extraída.

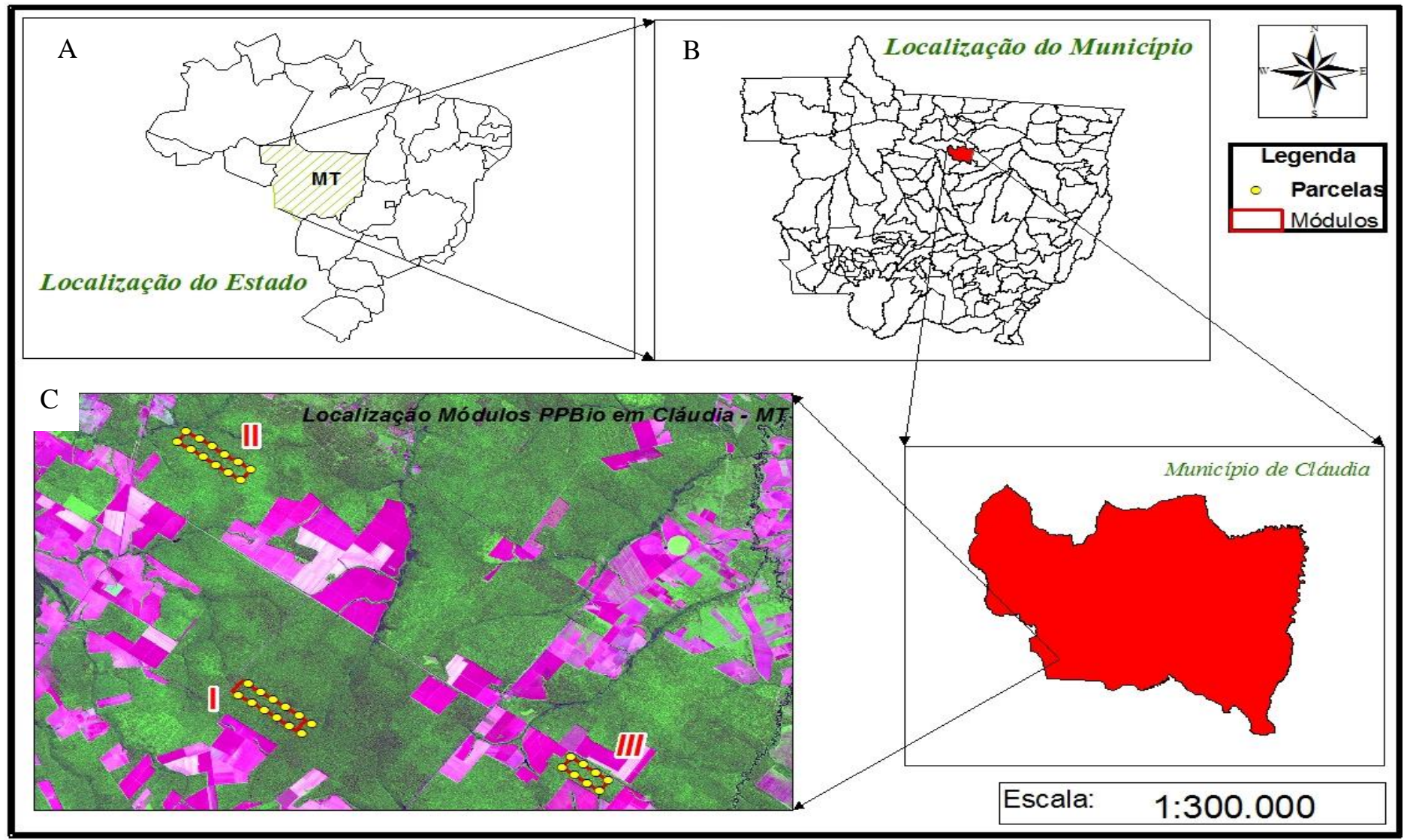

Figura1. Localização do estado de Mato Grosso no Brasil (A). Localização do município de Claudia no estado de Mato Grosso, Brasil (B). Localização da área amostral no município de Claudia, o retângulo vermelho representa os módulos e os círculos amarelos as parcelas permanentes (C). (Imagem criada por Everton José de Almeida programa Arcgis).

# 3. MATERIAL E MÉTODOS

## 3.1 Delineamento amostral

Foram instalados dois módulos na Fazenda Continental, e um terceiro na Fazenda Iracema, ambas pertencentes ao município de Cláudia. Os módulos instalados na Fazenda Continental são compostos por 2 trilhas de 5 km, espaçadas entre si por 1 km, formando um retângulo de 5 km² (Figura 1- A). O módulo instalado na Fazenda Iracema é composto por 2 trilhas de 3 km, espaçadas entre si por 1 km, formando um retângulo de 3 km² (Figura 1- B). Este delineamento foi desenvolvido com base em módulos do sistema RAPELD (MAGNUSSON et al, 2005), que permite inventários rápidos (RAP) e de longa duração (PELD), adotado pelo Programa de Pesquisa em Biodiversidade (PPBio-MCT) do Ministério de Ciência e Tecnologia do Brasil. A cada quilômetro ao longo da trilha, foi instalada uma parcela permanente de 250 m, resultando em 32 parcelas nas quais foram realizados os levantamentos biológicos padronizados, seguindo protocolo de coletas do PPBio (MAGNUSSON; MARTINS, 2005). A distribuição sistemática de parcelas na paisagem permite estimativas não tendenciosas da distribuição, abundância e biomassa das espécies em cada sítio, e comparações biogeográficas entre sítios. Todas as parcelas amostrais seguiam a curva de nível do terreno a fim de minimizar a variação interna das características do solo e topografia (MAGNUSSON et al, 2005). Foram instaladas ainda sete parcelas ripárias de 250 m, que definimos como hábitat terrestre adjacente a cursos d'água, sendo três delas no módulo 1 e duas nos módulos 2 e 3.

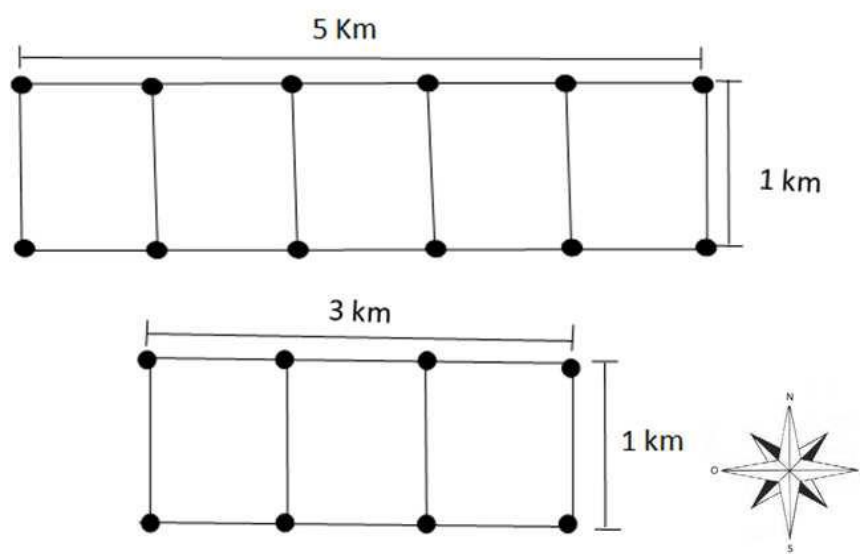

Figura 2: Esquema dos módulos 1 e 2 (A), e 3 (B), implantados nas três áreas de estudo, localizados no município de Cláudia, norte do estado de Mato Grosso.

### 3.2 Inventário de herbáceas

A comunidade herbácea inclusa na amostragem foram apenas espécies obrigatoriamente terrestres e as hemiepífitas (POULSEN, 1996). A amostragem foi feita entre setembro de 2010 e setembro de 2011. Dentro de cada parcela de 250 m x 2 m todos os indivíduos herbáceos maiores que 5 cm de altura foram contados, medidos e identificados. Algumas espécies ocorrem em grande densidade em uma pequena área, dificultando a diferenciação dos indivíduos, que nesses casos aparentam propagar-se de forma vegetativa. Para essas espécies, cada aglomerado de indivíduos foi considerado como um único indivíduo.

As identificações foram baseadas no guia de campo da Flora da Reserva Ducke (RIBEIRO et al., 1999), Guia de samambaias e licófitas da REBIO Uatumã (ZUQUIM et al., 2008), Guia de marantáceas da Reserva Ducke e da Reserva Biológica do Uatumã (COSTA et al., 2008a), Guia de Zingiberales dos sítios PPBio na Amazônia Ocidental brasileira (COSTA et al., 2011) e auxílio de bibliografias especializadas (KRAMER, 1957; ASSIS; SALINO, 2011; WINTER et al., 2011). A identificação das pteridófitas foram confirmadas pelo Dr. Alexandre Salino, taxonomista de pteridófitas da Universidade Federal de Minas Gerais e pelo Ms. Francisco de Paula Athayde Filho, professor assistente da Universidade do Estado de Mato Grosso, campus Nova Xavantina.

# 4. VARIÁVEIS AMBIENTAIS E DE PERTURBAÇÃO

### 4.1 Coletas de solo

As coletas de solo foram realizadas em seis pontos de cada uma das 39 parcelas, em uma profundidade de 0-10 cm, em 6 pontos a cada 50 m ao longo de cada parcela e homogeneizadas para gerar uma única amostra composta por parcela. O solo foi coletado com o auxílio de um trado holandês. Após secas ao ar livre as amostras foram destorroadas e peneiradas (peneira de 2 mm). As análises incluíram textura (percentual de silte, argila e areia), macro e micro nutrientes, pH e matéria orgânica, e foram baseadas nos métodos aplicados pela Empresa Brasileira de Pesquisa Agropecuária (EMBRAPA, 1999), no laboratório MT Solos - Análises Agronômicas S/C Ltda, Sorriso, MT.

## 4.2 Abertura de dossel

Para estimar a abertura de dossel foi utilizado um esferodensiômetro côncavo, (Concave – Model 1C – Robert E. Lemonn, Forest Densiometers, 5733 SE Cornell Dr. Bartlsville, OK, USA). Em cada parcela foram realizados registros de abertura de dossel a cada 50 m. Cada registro foi composto de quatro leituras (norte, sul, leste e oeste), sumarizadas como uma média. As leituras foram realizadas entre os horários de 9:00 h e 14:00 h.

## 4.3 Distância do curso d'água

A distância aproximada entre as parcelas e os igarapés presentes nos módulos foi medida com o auxílio do programa DIVA GIS 723. Foi aferida a menor distância (em linha reta), em metros das coordenadas geográficas das parcelas até os igarapés mais próximos encontrados na área de cada módulo.

## 4.4 Impactos do corte de madeira

O impacto causado pelo corte de madeira foi representado pelo número de árvores cortadas dentro de cada parcela. Os vestígios das árvores cortadas foram contados em uma área de 10 m para cada lado da linha central da parcela, e ao longo de todo o seu comprimento (250 m).

## 4.5 Análises dos dados

A composição das espécies de ervas foi ordenada através de Escalonamento Multidimensional Não-Métrico (NMDS) para os dados de abundância (quantitativos). O NMDS reduz as informações de um elevado número de atributos (espécies), até um pequeno número de variáveis ou eixos, que ordenam os pontos amostrais baseados numa distância estatística ajustada a partir das distâncias originais (MCCUNNE; GRACE, 2002). O NMDS não preserva as distâncias da matriz original, preserva apenas a relação de ordenação entre os objetos, permite o uso de qualquer matriz de distância e que o número de dimensões, para a ampliação do objeto, seja escolhido a priori (LEGENDRE; LEGENDRE, 1998). De acordo com Faith et. al (1987), NMDS foi considerado com um teste bastante robusto na identificação de padrões em ecologia da comunidade.

A dissimilaridade entre as unidades amostrais foi calculada usando a matriz de associação Bray-Curtis (MCCUNE; GRACE, 2002) para os dados previamente padronizados por parcela pela divisão pela soma de espécies. O resultado dessa padronização expressa a proporção de cada espécie em relação à abundância total de indivíduos por parcela.

Para testar a correlação entre as variáveis ambientais foi realizado o teste de Correlação de Spearman, que permite medir a intensidade da relação entre as variáveis ordinais e não exige pressuposto de distribuição normal (LIRA; CHAVES NETO, 2006). As variáveis testadas na Correlação de Spearman foram: altitude, abertura de dossel, serrapilheira, matéria orgânica, distância do curso d'água mais próximo, porcentagem de areia, de silte e de areia no solo, número total de árvores cortada, tempo de manejo e soma de bases (potássio, magnésio, e cálcio). As variáveis que apresentaram correlação superior a 0,4 foram excluídas nas análises.

As relações entre variáveis ambientais (variáveis independentes) e a composição da comunidade herbácea, expressas por um eixo de NMDS foram analisadas por regressões múltiplas com os seguintes modelos: (I) modelo simples: NMDS= constante + abertura de dossel + número de árvores cortadas + distância do curso d'água + teor de areia no solo; (II) modelo com interação: NMDS= constante + abertura de dossel + número de árvores cortadas + distância do curso d'água + teor de areia no solo + distância do curso d'água*abertura de dossel+ distância do curso d'água*teor de areia.

Para avaliar a relação entre as variáveis ambientais (variáveis independentes) e a riqueza das espécies, foram realizadas regressões múltiplas com os seguintes modelos: (I) modelo simples: Riqueza= constante + abertura de dossel + número de árvores cortadas + distância do curso d'água + teor de areia no solo; (II) modelo com interação: Riqueza= constante + abertura de dossel + número de árvores cortadas + distância do curso d'água + teor de areia no solo + distância do curso d'água*abertura de dossel+ distância do curso d'água*teor de areia.

Todas as análises estatísticas foram realizadas no programa R (R Development Core Team, 2011).

# 5. RESULTADOS

## 5.1 Variáveis ambientais e de perturbações

A Tabela 1 mostra as correlações entre as variáveis ambientais e de perturbação. O teor de silte assim como o teor de argila, soma de bases, fertilidade (matéria orgânica) e altitude, esteve altamente correlacionado (de forma negativa) com o teor de areia no solo.

Portanto foram excluídas das análises. A serapilheira esteve correlacionada com teor de silte, soma de bases, fertilidade do solo, abertura de dossel, e também não foi incluída nas análises. O tempo de manejo esteve correlacionado com teor de areia e com soma de bases (de forma negativa), sendo assim excluído das analises.

As variáveis que foram incluídas nas análises foram apenas teor de areia no solo, abertura de dossel, número de árvores cortadas e distância do curso d'água.

Tabela 1 - Resultado da Correlação de Spearman entre as variáveis ambientais analisadas nas parcelas:Altitude, Abertura de dossel (Dossel), Serrapilheira, Matéria orgânica (Mat. orgânica), Teor de areia no solo (Areia), Teor de silte no solo (Silte), Teor de Argila no solo (Argila), número de árvores cortadas (N. árvores cortadas), Tempo pós-manejo (Tempo manejo), Distância do Curso d'água (Dist. curso d'água) e Soma de bases (S. bases).

| | Dossel | Serrapilheira | Dist curso d'água | Altitude | Areia | Silte | Argila | Mat. orgânica | Soma bases | N. árvores cortadas | Tempo manejo |
|---|---|---|---|---|---|---|---|---|---|---|---|
| **Dossel** | 1 | | | | | | | | | | |
| Serrapilheira | 0.71 | 1 | | | | | | | | | |
| **Dist. curso d'água** | -0.27 | -0.36 | 1 | | | | | | | | |
| Altitude | 0.19 | 0.04 | 0.52 | 1 | | | | | | | |
| **Areia** | -0.29 | -0.23 | -0.26 | -0.64 | 1 | | | | | | |
| Silte | 0.47 | 0.44 | 0.07 | 0.42 | -0.75 | 1 | | | | | |
| Argila | 0.24 | 0.14 | 0.31 | 0.63 | -0.97 | 0.66 | 1 | | | | |
| Mat. Orgânica | 0.56 | 0.58 | -0.09 | 0.25 | -0.49 | 0.51 | 0.47 | 1 | | | |
| S. bases | 0.41 | 0.44 | 0.14 | 0.57 | -0.69 | 0.48 | 0.67 | 0.55 | 1 | | |
| **N. árvores Cortadas** | -0.38 | -0.33 | 0.26 | 0.02 | 0.1 | -0.31 | -0.08 | -0.35 | 0.03 | 1 | |
| Tempo manejo | -0.25 | -0.33 | 0.05 | -0.25 | 0.52 | -0.29 | -0.45 | -0.15 | -0.55 | -0.19 | 1 |

## 5.2 Composição da comunidade herbácea

Foram inventariados um total de 7.965 indivíduos, distribuídos em 70 espécies, 38 gêneros e 22 famílias (Tabela 2). Nas parcelas de interflúvio foram encontrados 3.952 indivíduos e nas parcelas ripárias foram encontrados 4.013 indivíduos, sendo 759 hemiepífitas e 7.208 terrestres.

As famílias mais ricas em espécies foram Cyperaceae e Pteridaceae com 10 espécies cada uma. As espécies *Diplasia karataefolia* e *Scleria secans* (Cyperaceae), *Adiantum lucidum* e *Adiantum obliquum* (Pteridaceae), *Triplophyllum hirsutum* (Tectariaceae), *Trichomanes pinnatum* (Hymenophyllaceae), *Ischnosiphon* sp1 (Marantaceae) e *Streptogyna americana* e *Pariana* sp1 (Poaceae) ocorreram em todas as parcelas.

Tabela 2: Famílias e espécies herbáceas, hábito, número de indivíduos encontrados nas parcelas de interflúvio (32 parcelas) e ripárias (7 parcelas) (abundância) e número total de indivíduos encontrados em 39 Parcelas das três áreas localizadas no município de Cláudia, norte do estado de Mato Grosso.

| Familias/espécies | Hábito | Parcelas de interflúvio (abundância) | Parcelas ripárias (abundância) | N° de indivíduos |
|---|---|---|---|---|
| (Angiospermae) | | | | |
| **ARACEAE** | | | | |
| *Anaphyllopsis cururuana* A. Hay | Terrestre | 0 | 5 | 5 |
| *Montrichardia arborescens* Schott | Terrestre | 0 | 50 | 50 |
| *Philodendron* sp1 | Hemiepífita | 2 | 4 | 6 |
| *Philodendron* sp2 | Hemiepífita | 0 | 22 | 22 |
| *Philodendron* sp3 | Hemiepífita | 0 | 3 | 3 |
| *Philodendron* sp4 | Hemiepífita | 0 | 1 | 1 |
| **BROMELIACEAE** | | | | |
| Não identificada | Terrestre | 6 | 0 | 6 |
| **COSTACEAE** | | | | |
| *Costus arabicus* L. | Terrestre | 0 | 5 | 5 |
| *Costus* sp. | Terrestre | 0 | 7 | 7 |
| **CYPERACEAE** | | | | |
| *Calyptrocarya* sp1 | Terrestre | 2 | 0 | 2 |
| *Calyptrocarya* sp2 | Terrestre | 14 | 220 | 234 |
| *Calyptrocarya* sp3 | Terrestre | 1 | 0 | 1 |
| *Calyptrocarya* sp4 | Terrestre | 2 | 18 | 20 |
| *Diplasia karataefolia* Rich. | Terrestre | 80 | 11 | 91 |
| *Hypolytrum* sp. | Terrestre | 0 | 8 | 8 |
| *Scleria secans* Urb. | Terrestre | 27 | 19 | 46 |
| *Scleria* sp1 | Terrestre | 3 | 0 | 3 |
| *Scleria* sp2 | Terrestre | 1 | 0 | 1 |

continuação da tabela 2

| Familias/espécies | Hábito | Parcelas de interflúvio (abundância) | Parcelas ripárias (abundância) | N° de indivíduos |
|---|---|---|---|---|
| *Scleria* sp3 | Terrestre | 0 | 44 | 44 |
| **HELICONIACEAE** | | | | |
| *Heliconia* sp | Terrestre | 2 | 4 | 6 |
| **MARANTACEAE** | | | | |
| *Calathea zingiberina* Körn. | Terrestre | 29 | 0 | 29 |
| *Calathea* sp1 | Terrestre | 1 | 0 | 1 |
| *Ischnosiphon* sp1 | Terrestre | 187 | 144 | 331 |
| *Ischnosiphon* sp2 | Terrestre | 26 | 16 | 42 |
| *Monotagma densiflorum* K. Schum. | Terrestre | 0 | 2 | 2 |
| *Monotagma plurispicatum* K. Schum. | Terrestre | 25 | 162 | 187 |
| **MORACEAE** | | | | |
| *Dorstenia* sp | Terrestre | 0 | 17 | 17 |
| **POACEAE** | | | | |
| Aff. *Lasiacis* | Terrestre | 1 | 0 | 1 |
| *Bambusa* sp. | Terrestre | 27 | 0 | 27 |
| *Ichnanthus panicoides* P. Beauv. | Terrestre | 1 | 0 | 1 |
| *Olyra latifolia* L. | Terrestre | 80 | 0 | 80 |
| *Olyra* sp3 | Terrestre | 1 | 0 | 1 |
| *Pariana* sp1 | Terrestre | 307 | 559 | 866 |
| *Pariana* sp2 | Terrestre | 37 | 0 | 37 |
| *Pariana* sp3 | Terrestre | 11 | 70 | 81 |
| *Pariana* sp4 | Terrestre | 0 | 129 | 129 |
| *Streptogyna americana* C. E. Hubb. | Terrestre | 968 | 6 | 974 |

continuação da tabela 2

| Familias/espécies | Hábito | Parcelas de interflúvio (abundância) | Parcelas ripárias (abundância) | N° de indivíduos |
|---|---|---|---|---|
| **RAPATEACEAE** | | | | |
| *Rapatea paludosa* Aubl. | Terrestre | 0 | 186 | 186 |
| **STRELITZIACEAE** | | | | |
| *Phenakospermum guianensis* Aubl. (Pteridófitas) | Terrestre | 62 | 31 | 93 |
| **CYATHEACEAE** | | | | |
| *Cyathea pungens* (Willd.) Domin. | Terrestre | 0 | 56 | 56 |
| **DRYOPTERIDACEAE** | | | | |
| *Cyclodium meniscioides* (Willd.) C. Presl | Terrestre | 1 | 58 | 59 |
| *Mickelia guianensis* (Aubl.) R. C. Moran, Labiak & Sundue. | Hemiepífita | 0 | 339 | 339 |
| *Polybotrya sessilisora* R.C. Moran | Hemiepífita | 0 | 69 | 69 |
| **HYMENOPHYLLACEAE** | | | | |
| *Trichomanes ankersii* C. Parker ex Hook. & Grev. | Hemiepífita | 0 | 38 | 38 |
| *Trichomanes cristatum* Kaulf. | Terrestre | 0 | 16 | 16 |
| *Trichomanes pinnatum* Hedw. | Terrestre | 142 | 640 | 782 |
| *Trichomanes vittaria* DC. ex Poir. | Terrestre | 194 | 13 | 207 |
| **LINDSAEACEAE** | | | | |
| *Lindsaea divaricata* Klotzsch. | Terrestre | 0 | 22 | 22 |
| *Lindsaea guianensis* (Aubl.) Dryand. | Terrestre | 0 | 6 | 6 |
| *Lindsaea lancea* (L.) Bedd. var. falcata (Dryand.) Rosenst. | Terrestre | 0 | 59 | 59 |

continuação da tabela 2

| Familias/espécies | Hábito | Parcelas de interflúvio (abundância) | Parcelas ripárias (abundância) | N° de indivíduos |
|---|---|---|---|---|
| *Lindsaea lancea* (L.) Bedd. var. lancea Christ. | Terrestre | 10 | 80 | 90 |
| *Lindsaea quadrangularis* Raddi subsp. quadrangularis. | Terrestre | 3 | 28 | 31 |
| **LOMARIOPSIDACEAE** | | | | |
| *Lomariopsis japurensis* (C.Martius) J.Sm. | Hemiepífita | 218 | 63 | 281 |
| **MARATTIACEAE** | | | | |
| *Danaea leprieurii* Kunze. | Terrestre | 0 | 229 | 229 |
| **METAXYACEAE** | | | | |
| *Metaxya rostrata* C. Presl | Terrestre | 0 | 402 | 402 |
| **PTERIDACEAE** | | | | |
| *Adiantum cinnamomeum* Lellinger & J. Prado. | Terrestre | 6 | 0 | 6 |
| *Adiantum diogoanum* Glaziou. | Terrestre | 27 | 0 | 27 |
| *Adiantum* cf *diogoanum* | Terrestre | 1 | 0 | 1 |
| *Adiantum dolosum* Kunze. | Terrestre | 212 | 9 | 221 |
| *Adiantum glaucescens* Klotzsch. | Terrestre | 52 | 0 | 52 |
| *Adiantum lucidum* (Cav.) Sw. | Terrestre | 650 | 32 | 682 |
| *Adiantum obliquum* Willd. | Terrestre | 206 | 4 | 210 |
| *Adiantum* cf. *obliquum* | Terrestre | 3 | 0 | 3 |
| *Adiantum tetraphyllum* Humb. & Bonpl. ex Willd. | Terrestre | 6 | 0 | 6 |
| *Adiantum tomentosum* Klotzsch. | Terrestre | 0 | 33 | 33 |
| **SACCOLOMATACEAE** | | | | |
| *Saccoloma inaequale* (Kunze) Mett | Terrestre | 0 | 26 | 26 |

continuação tabela 2

| Familias/espécies | Hábito | Parcelas de interflúvio (abundância) | Parcelas ripárias (abundância) | N° de indivíduos |
|---|---|---|---|---|
| **SCHIZAEACEAE** | | | | |
| *Schizaea elegans* (Vahl) Sw. | Terrestre | 1 | 0 | 1 |
| **SELAGINELLACEAE** | | | | |
| *Selaginella breynii* Spring. | Terrestre | 0 | 2 | 2 |
| **TECTARIACEAE** | | | | |
| *Triplophyllum hirsutum* (Holttum) J. Prado & R.C. Moran. | Terrestre | 308 | 40 | 348 |
| **THELYPTERIDACEAE** | | | | |
| *Thelypteris leprieurii* (Hook.) R.M. Tryon. | Terrestre | 0 | 17 | 17 |

## 5.3 Relações entre variável dependente X variável independente

A ordenação da composição quantitativa por NMDS captou 65 % da variação com 1eixo (Stress=0,26). A regressão múltipla com modelo simples sugeriu que a composição da comunidade herbácea foi influenciada apenas pela distância do curso d'água (Tabela 3) e que as demais variáveis não afetaram a composição (Figura 3).

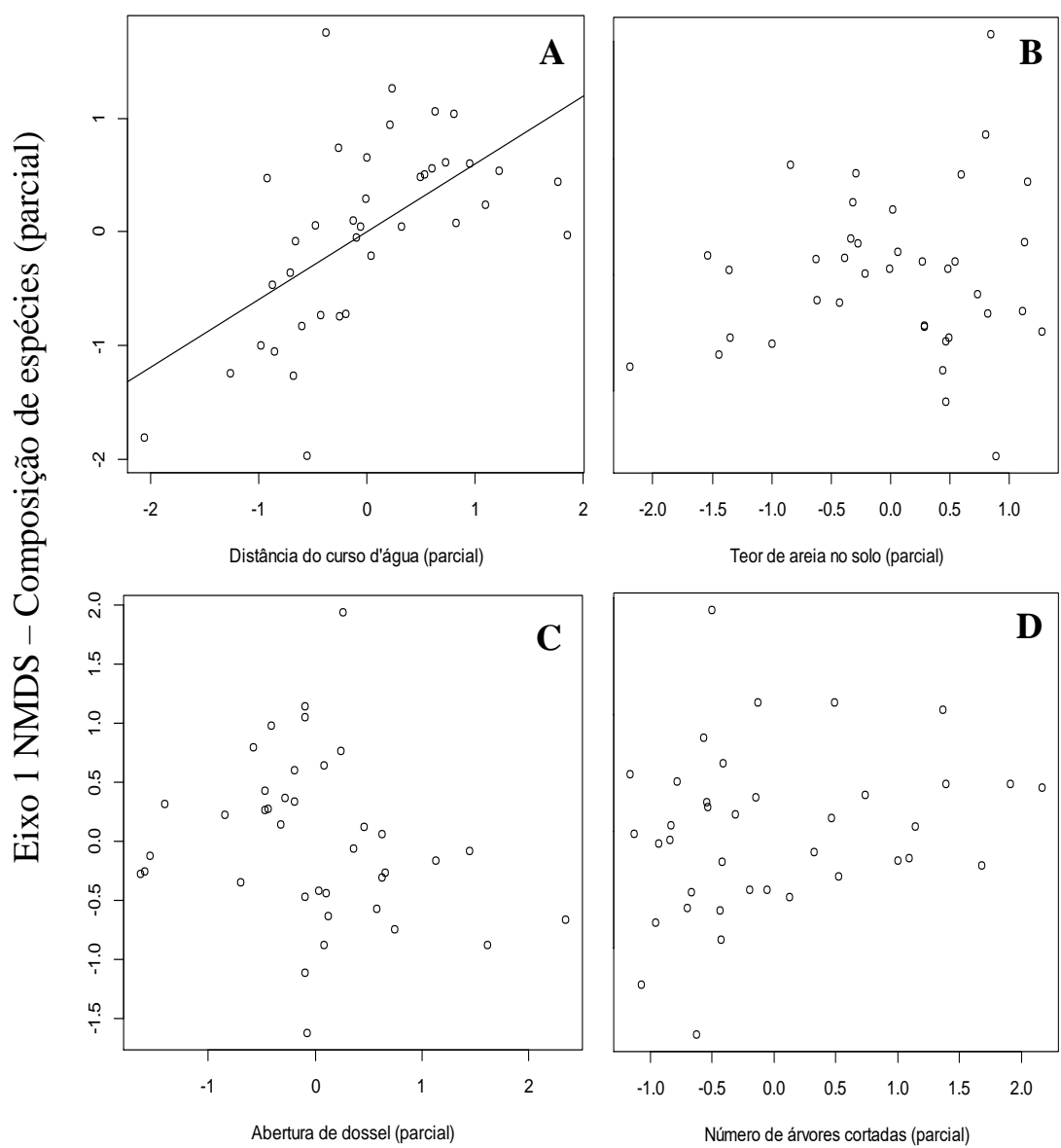

Figura 3. Relações entre a composição da comunidade herbácea resumida em um eixo do NMDS e as variáveis ambientais preditoras. A) Distância do curso d'água, B) Teor de areia no solo ,C) Abertura de dossel e D) Número de árvores cortadas, em 39 parcelas das três áreas localizadas no município de Cláudia, norte do estado de Mato Grosso.

Tabela 3: Resultado das regressões múltiplas entre composição e riqueza de espécies de ervas
(valor dos coeficientes parciais da regressão, entre parênteses os valores das probabilidades)
e as variáveis preditoras para 39 parcelas das três áreas localizadas no município de Cláudia,

| | Nº árvores cortadas | Distância do curso d'água | Abertura de dossel | Teor de areia | Dist.do curso d'água+abertura de dossel | Dist.do curso d'água+teor de areia | $R^2$ |
|---|---|---|---|---|---|---|---|
| Composição de espécies (modelo simples) | -0.17 (0.189) | -0.60 (0.000) | 0.18 (0.205) | -0.10 (0.486) | - | - | 0.53 |
| Composição de espécies (modelo com interação) | 0.10 (0.957) | -0.80 (0.000) | -0.27 (0.143) | -0.45 (0.008) | -0.49 (0.003) | -0.48 (0.008) | 0.66 |
| Riqueza de espécies (modelo simples) | -0.19 (0.178) | -0.35 (0.026) | 0.38 (0.015) | -0.04 (0.809) | - | - | 0.47 |
| Riqueza de espécies (modelo com interação) | - 0.06 (0.696) | -0.50 (0.003) | 0.04 (0.856) | -0.31 (0.102) | -0.39 (0.035) | -0.34 (0.090) | 0.55 |

norte do estado de Mato Grosso.

Entretanto, os gráficos das parciais mostram relações não lineares que sugerem interações entre alguns fatores. O modelo com interações detectou interações significativas entre a distância do curso d'água e o teor de areia (Tabela 3) e entre a distância do curso d'água e abertura de dossel (Figura 4). O teor de areia no solo influenciou a composição de espécies dependendo da distancia do curso d'água. Se o local é muito arenoso e próximo de um curso d'água a composição será diferente de locais arenosos e mais distantes da água. Áreas muito argilosas não existem na beira dos cursos d'água na região de estudo e, portanto a composição associada a elas não depende da associação com o curso d'água. O mesmo ocorre para o efeito da luminosidade, em que áreas mais iluminadas (>25% abertura dossel) apresentam composição diferente dependendo da distância do curso d'água, enquanto áreas menos iluminadas têm composição mais constante. O número de árvores cortadas não afetou a composição de espécies herbáceas em nenhum dos modelos considerados (Figuras 3C e 4D).

A ordenação direta ao longo da distância do curso d'água mostra um grupo de 17 espécies (35,4 % da comunidade) claramente restrito aos ambientes ripários e portanto mais úmidos. Um grupo de 13 espécies esteve restrito aos ambientes não-ripários distantes dos cursos d'água e cerca de 18 espécies compartilharam ambos os ambientes. A maior parte das espécies que ocorreu nos ambientes ripários e não-ripários esteve restrita às áreas menos distantes dos cursos d'água (até ~100 m) ou diminuiu em abundância conforme esta distância aumentou. Apenas *Adiantum lucidum* e *Ischnosiphon* sp1, que ocorrem nos dois ambientes, aumentaram sua abundancia nas áreas mais distantes dos cursos d'água (Figura 5).

A distribuição das espécies ao longo do gradiente de teor de areia não apresentou um padrão tão bem definido (Figura 6). O grupo de espécies associado às áreas mais argilosas é menos variável, ao contrário do grupo associado às áreas mais arenosas, que apresentou variações, tal como previsto pela interação entre o teor de areia e a distância do curso d'água.

Da mesma forma que acontece com o teor de areia, a ordenação direta das parcelas ao longo do gradiente de abertura do dossel não mostrou um padrão tão bem definido, mesmo assim, nota-se espécies com preferência por locais com mais luz e outras com menos luz (Figura 7). Não podemos esquecer que a distância do igarapé interage com a abertura de dossel, e assim a composição das áreas com maior entrada de luz é mais variável.

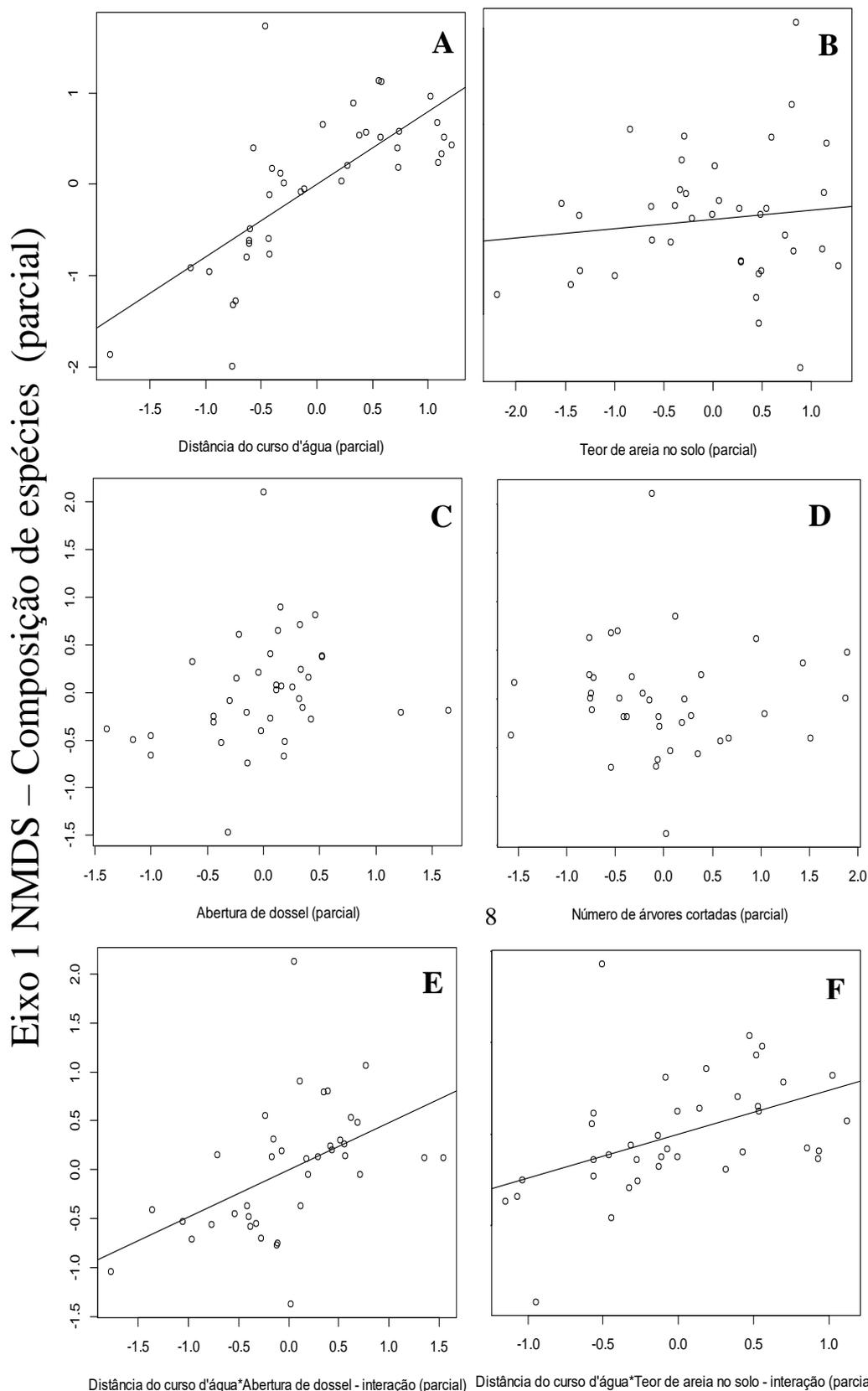

Figura 4. Relações entre a composição da comunidade herbácea resumida em um eixo de NMDS e interações entre as variáveis ambientais preditoras. A) Distância do curso d'água, B) Teor de areia no solo, C) Abertura de dossel, D) Número de árvores cortadas, E) Interação entre Distância do curso d'água e abertura de dossel e F) Interação entre Distância do curso d'água e teor de areia no solo em 39 parcelas das três áreas localizadas no município de Cláudia, norte do estado de Mato Grosso.

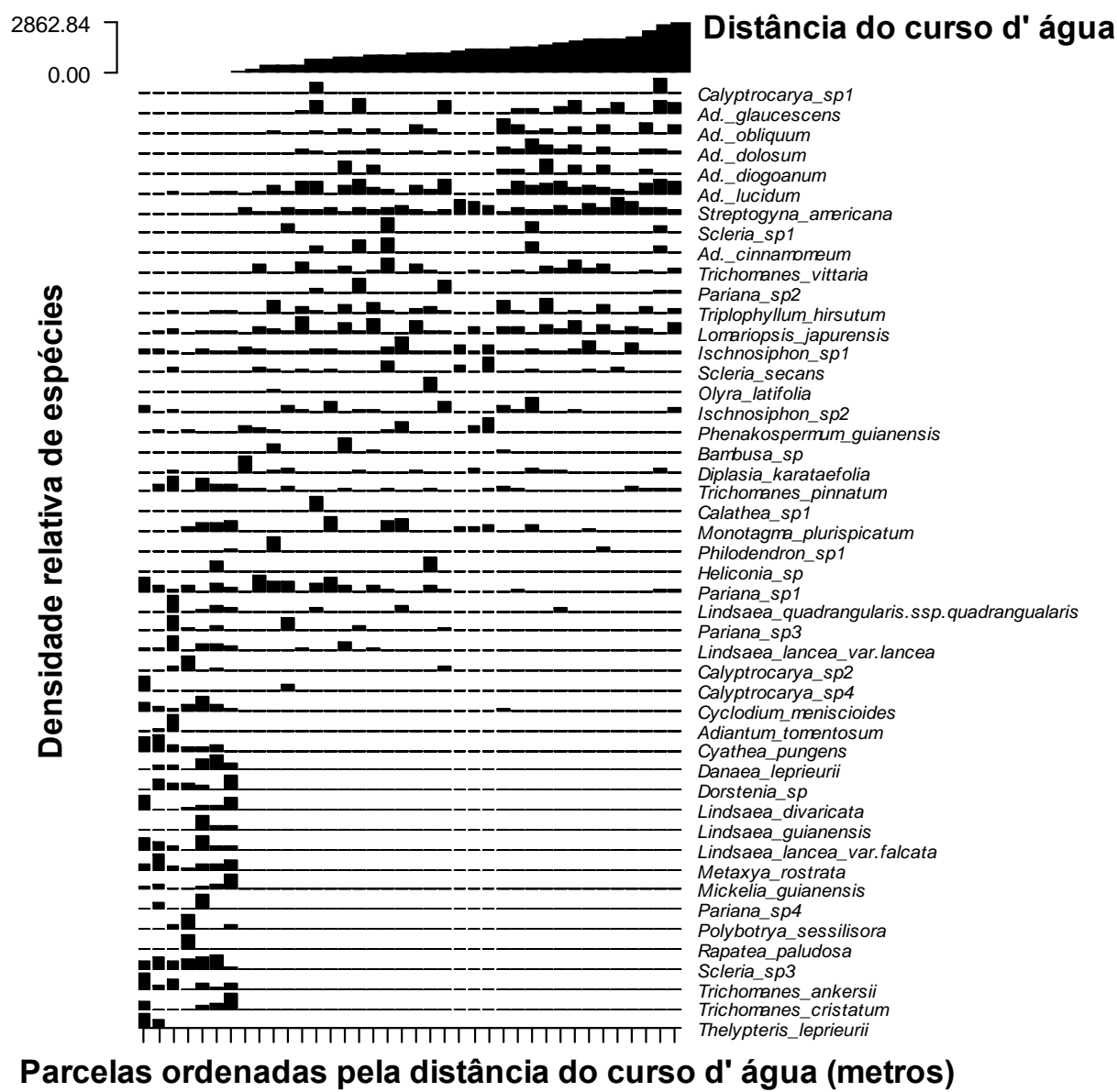

Figura 5. Distribuição das espécies no gradiente Distância do curso d'água, em 39 parcelas das três áreas localizadas no município de Cláudia, norte do estado de Mato Grosso. A ordem das espécies no gráfico foi construída em função dos gradientes originais. (As espécies que ocorreram uma única vez durante a amostragem não aparecem no gráfico).

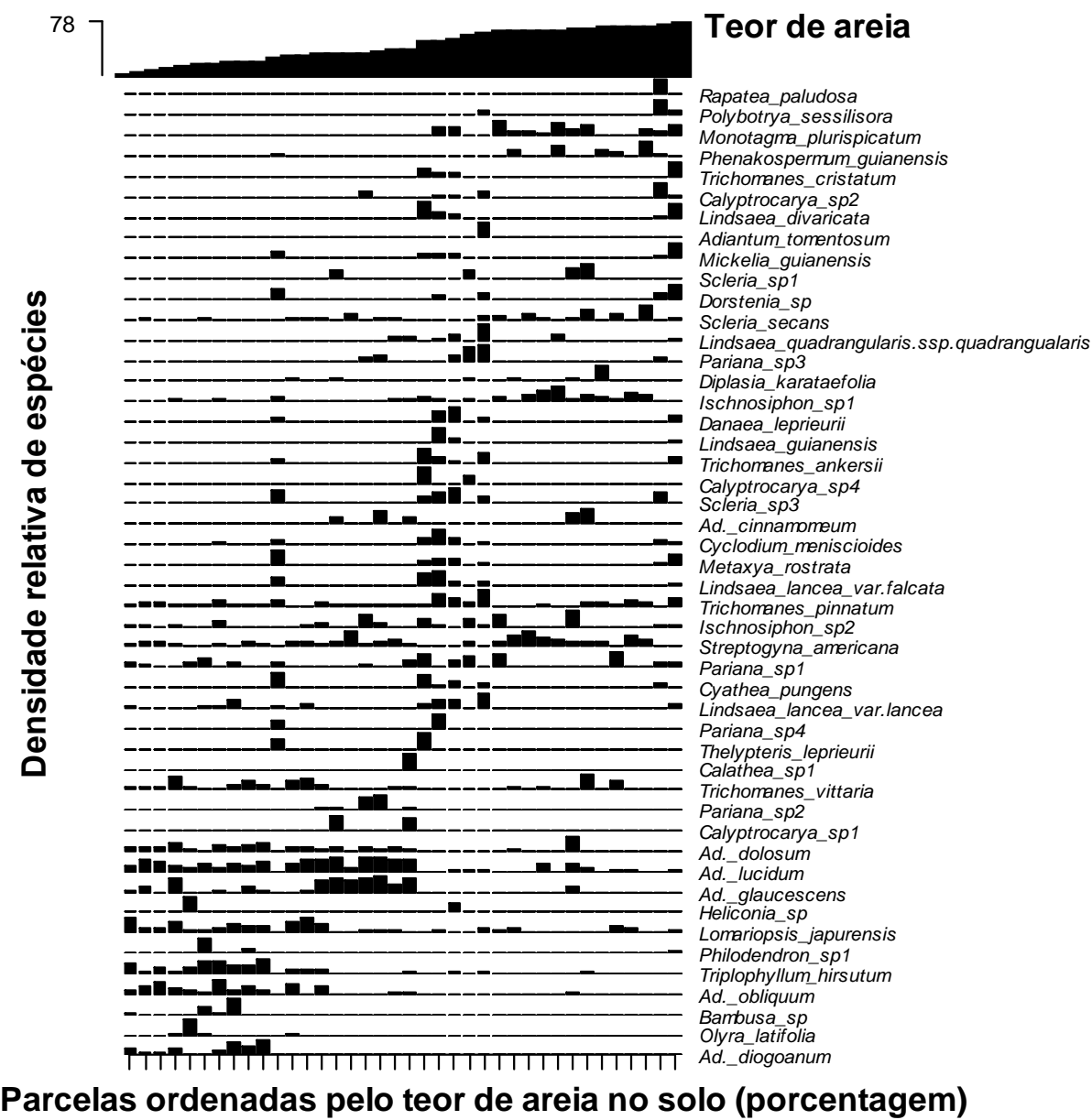

**Teor de areia**

78

**Densidade relativa de espécies**

Rapatea_paludosa
Polybotrya_sessilisora
Monotagma_plurispicatum
Phenakospermum_guianensis
Trichomanes_cristatum
Calyptrocarya_sp2
Lindsaea_divaricata
Adiantum_tomentosum
Mickelia_guianensis
Scleria_sp1
Dorstenia_sp
Scleria_secans
Lindsaea_quadrangularis.ssp.quadrangualaris
Pariana_sp3
Diplasia_karataefolia
Ischnosiphon_sp1
Danaea_leprieurii
Lindsaea_guianensis
Trichomanes_ankersii
Calyptrocarya_sp4
Scleria_sp3
Ad._cinnamomeum
Cyclodium_meniscioides
Metaxya_rostrata
Lindsaea_lancea_var.falcata
Trichomanes_pinnatum
Ischnosiphon_sp2
Streptogyna_americana
Pariana_sp1
Cyathea_pungens
Lindsaea_lancea_var.lancea
Pariana_sp4
Thelypteris_leprieurii
Calathea_sp1
Trichomanes_vittaria
Pariana_sp2
Calyptrocarya_sp1
Ad._dolosum
Ad._lucidum
Ad._glaucescens
Heliconia_sp
Lomariopsis_japurensis
Philodendron_sp1
Triplophyllum_hirsutum
Ad._obliquum
Bambusa_sp
Olyra_latifolia
Ad._diogoanum

**Parcelas ordenadas pelo teor de areia no solo (porcentagem)**

Figura 6. Distribuição das espécies ao longo do gradiente de teor de areia no solo, em 39 parcelas das três áreas localizadas no município de Cláudia, norte do estado de Mato Grosso. A ordem das espécies no gráfico foi construída em função dos gradientes originais. (As espécies que ocorreram uma única vez durante a amostragem não aparecem no gráfico).

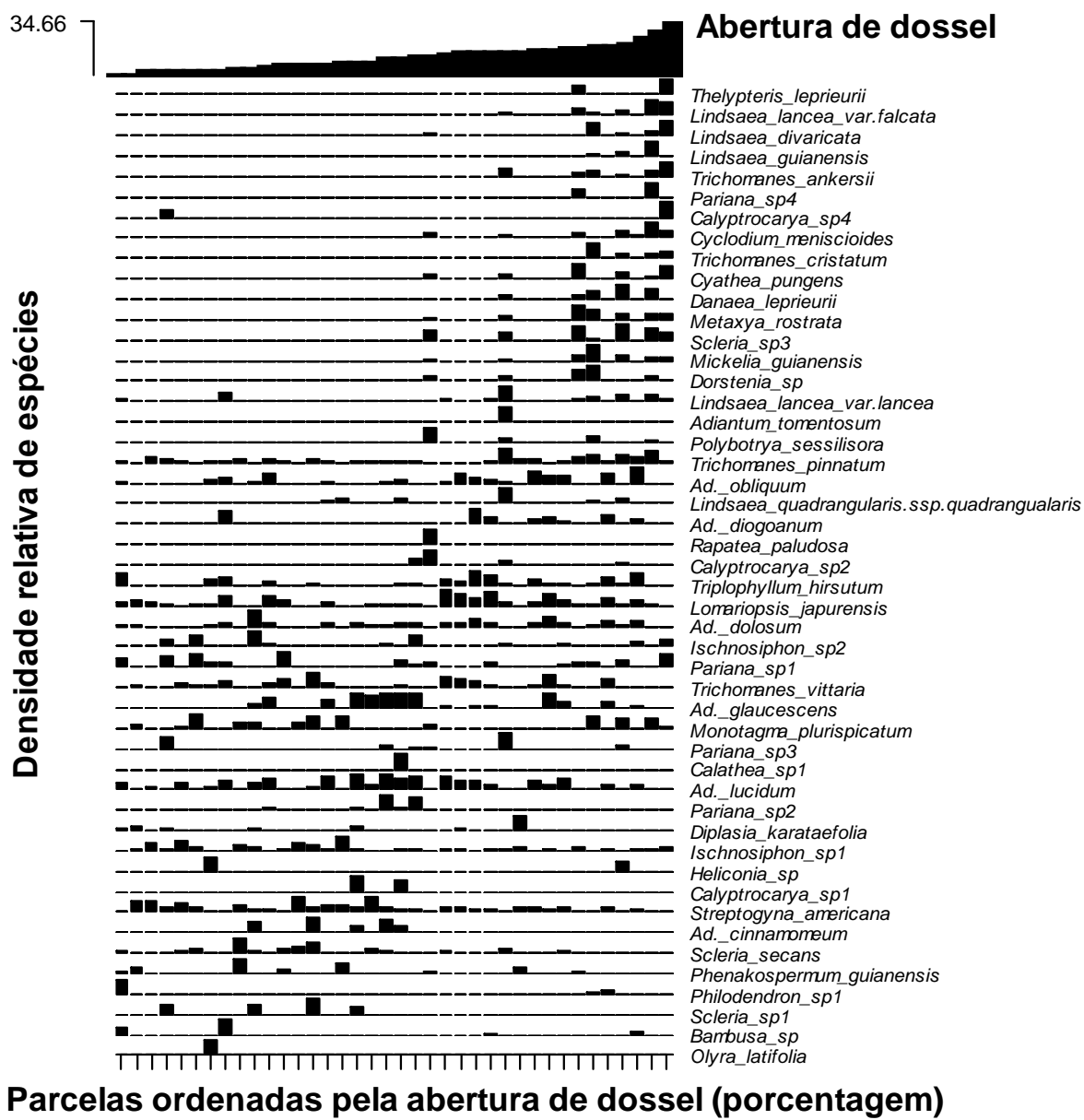

**Parcelas ordenadas pela abertura de dossel (porcentagem)**

Figura 7. Distribuição das espécies ao longo do gradiente de abertura do dossel, em 39 parcelas das três áreas localizadas no município de Cláudia, norte do estado de Mato Grosso. A ordem das espécies no gráfico foi construída em função dos gradientes originais. (As espécies que ocorreram uma única vez durante a amostragem não aparecem no gráfico).

### 5.4 Riqueza de espécies

O modelo simples de regressão múltipla multivariada (incluindo distância do curso d'água, abertura de dossel, teor de areia e número de árvores cortadas como variáveis independentes) mostra que a riqueza de espécies herbáceas foi influenciada pela distância do curso d'água (Tabela 3) e pela abertura de dossel, mas não teve influência do número de árvores cortadas e do teor de areia no solo (Figura 8).

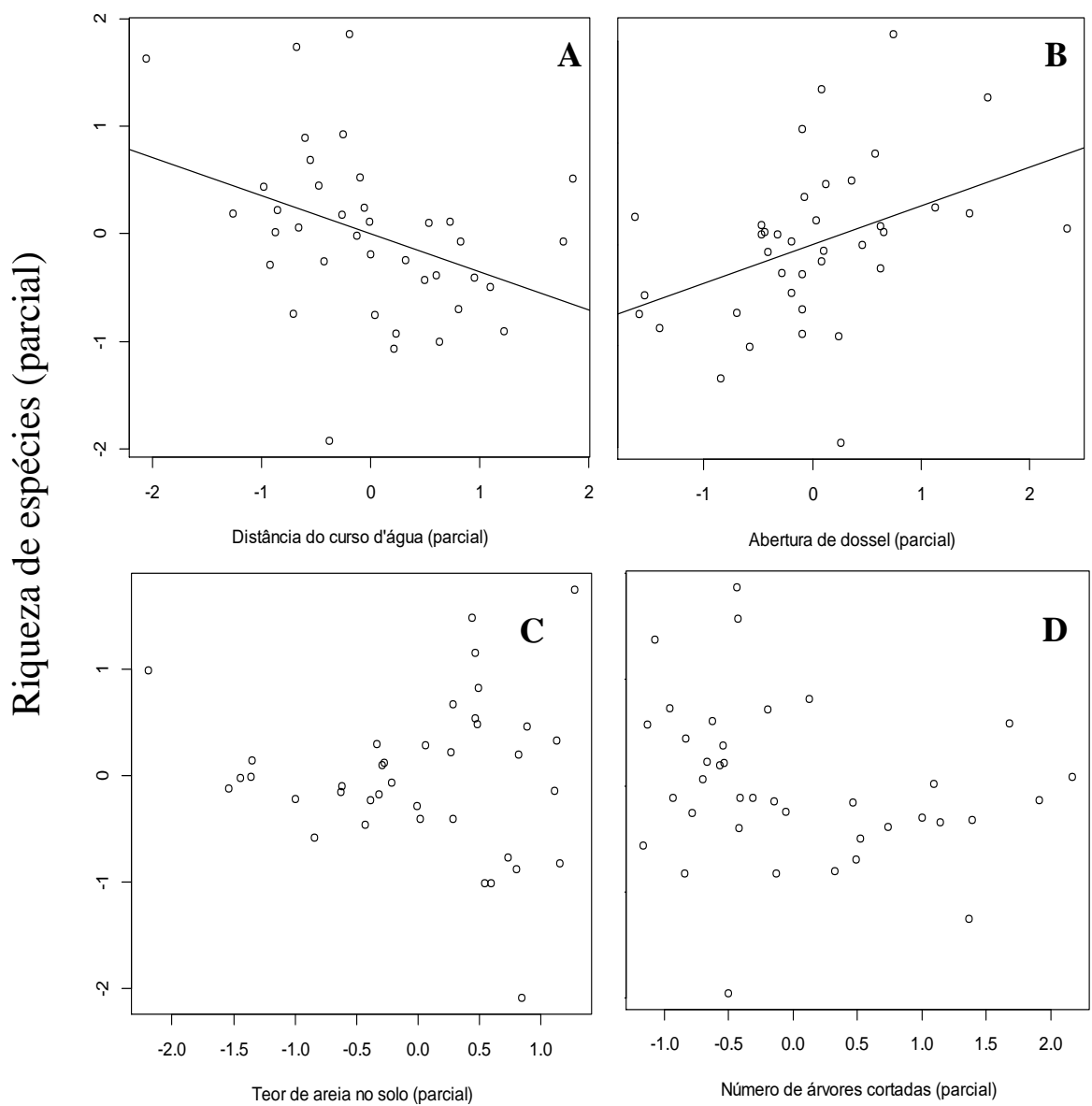

Figura 8. Relações entre a Riqueza de espécies herbácea e às variáveis ambientais preditoras. A) Distância do curso d'água, B) Abertura de dossel, C) Teor de areia no solo e D) Número de árvores cortadas, em 39 parcelas das três áreas localizadas no município de Cláudia, norte do estado de Mato Grosso.

Incluindo as mesmas interações consideradas na análise da composição de espécies (Distância do curso d'água*abertura de dossel; distância do curso d'água*teor de areia), observamos (Tabela 3) que distância do curso d'água também influência a maneira como a abertura de dossel pode estar determinando a riqueza de espécies (Figura 9). A riqueza de espécies diminui à medida que se vai distanciando do curso d'água. Em locais onde há maior entrada de luz a riqueza de espécies aumenta. A interação entre o teor de areia e a distancia do curso d'água foi marginalmente significativa (Tabela 3), havendo mais espécies em áreas arenosas mais próximas do curso d'água.

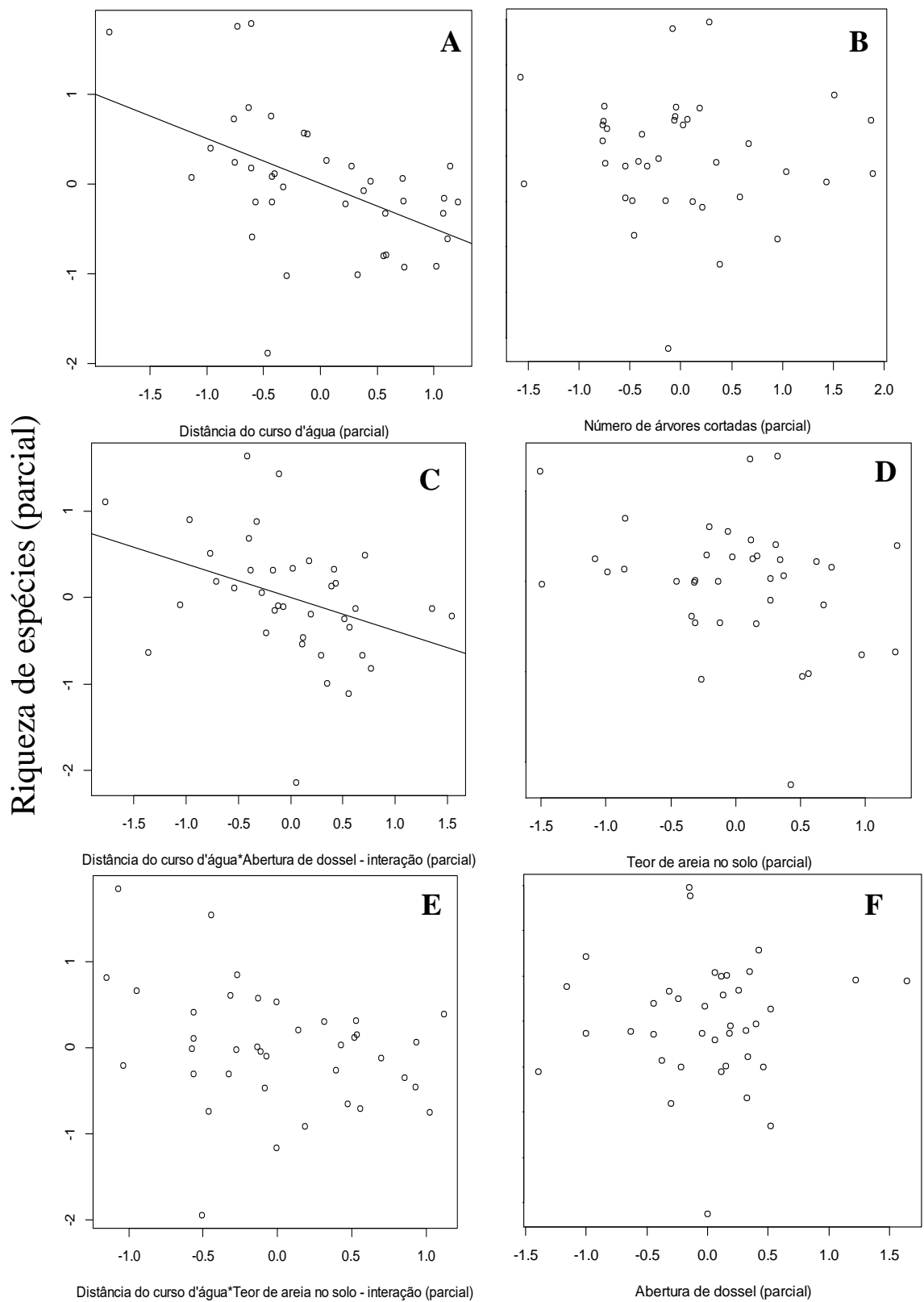

Figura 9. Relações entre a riqueza de espéceis e interações entre as variáveis ambientais preditoras. A) Distância do curso d'água, B) Número de árvores cortadas, C) Distância do curso d'água*abertura de dossel, D) Teor de areia no solo, E) Distância do curso d'água*teor de areia no solo e F)Abertura de dossel,em 39 parcelas das três áreas localizadas no município de Cláudia, norte do estado de Mato Grosso.

# 6. DISCUSSÃO

A variação na composição da comunidade herbácea estudada relacionou-se com o gradiente de distância do curso d'água, luminosidade e textura do solo evidenciados nas análises de NMDS e regressão múltipla. Houve uma substituição de espécies à medida que se distanciou do curso d'água, que se revelou o preditor mais importante da composição da comunidade herbácea neste estudo. As espécies de ervas se substituem ao longo do gradiente de distância do curso d'água, sugerindo que, para estre grupo, há oportunidade de segregação de nichos. Algumas espécies ocorreram exclusivamente próximas aos cursos dá água, sugerindo que as mesmas podem não sobreviver a condições de menor disponibilidade de água, enquanto *Adiantum glaucescens, Adiantum diogoanum, Adiantum cinnamomeum, Bambusa* sp, *Calyptrocarya* sp1, *Pariana* sp2, *Scleria* sp1, por exemplo ocorreram unicamente no interflúvio, área com menor disponibilidade de água. A umidade é apontada com um bom preditor da composição da comunidade vegetal (NAIMAN et al., 2000), pois afeta diretamente as condições de crescimento da vegetação em áreas ripárias (NAIMAN et al., 1998).

Estudos da comunidade arbórea mostraram uma forte restrição de certos grupos de espécies à áreas próximas a cursos d'água (ANDRADE et al., 2006; BERTANI et al., 2001; CAMPOS; LANDGRAF, 2001; FERREIRA; PAROLIN 2007; SALIS et al., 1994), mas a diferenciação da comunidade herbácea ao longo deste gradiente é pouco documentado. Um exemplo para Amazônia Central foi o estudo de Drucker et al. (2008) que observou diferenciação na composição de ervas ao longo do gradiente de distância do igarapé.

A extensão real da zona ripária, conforme percebida pelas plantas herbáceas foi de ~110m neste estudo. Resultado similar ao observado por Drucker et al. (2008) em uma área bem distante, na Amazônia central (~ 800km). As aves de sub-bosque estudadas na Amazônia central também apresentaram composicao diferenciada numa faixa de 100 a 140 m de distância dos cursos d'água (BUENO, 2011). Estes resultados congruentes permitem sugerir que esta faixa de ~100 m nas margens dos cursos d'água caracterizam a zona ripária do ponto de vista da biodiversidade.

A composição da comunidade também foi influenciada pela abertura do dossel. Muitas espécies estiveram associadas unicamente a ambientes mais úmidos e mais iluminados. De acordo com Hogan e Machado (2002), apenas 2% da luz solar incidente atinjem o solo de florestas tropicais, e a disponibilidade de luz é um recurso potencialmente limitante para espécies de sub-bosque (LITE et al., 2005; XIONG et al., 2003). Smith (1987) em seu estudo sobre ervas de sub-bosque no Panamá sugeriu que grande parte da variação de distribuição destas pode ser explicada por aberturas no dossel causadas por árvores caídas. Com a mortalidade de árvores de grande porte ocorre a invasão de ervas que colonizam com maior frequência esses ambientes, onde a entrada de luz é maior (NAIMAN et al., 2000). Ervas que se estabelecem em áreas abertas, apresentam altas taxas de crescimento sob condições de luminosidade elevada que lhes permitem competir por espaço e reproduzir-se rapidamente (SMITH, 1987).

Jones et. al (2006) encontrou efeito suave de abertura de dossel sobre a composição de pteridófitas da Amazônia. Um efeito sutil de abertura de dossel sobre a comunidade de pteridófitas em áreas amazônicas, também foi encontrado por Zuquim (2006), porém ao estudar ervas terrestres em áreas ripárias, Drucker (2008) não encontrou relação entre a abertura de dossel e a composição da comunidade. Apesar de o presente trabalho ter sido realizado sob condições similares ao de Drucker (2008), tanto a distância do curso d'água quanto a abertura do dossel foram suficientes para determinar diferenças na composição da comunidade.

O teor de areia no solo afetou a variação da comunidade herbácea dependendo da distância do igarapé, o que não foi observado em outros estudos de áreas amazônicas (COSTA et. al, 2005; COSTA et. al, 2008b; SOUSA, 2007; TUOMISTO et. al, 2003; ZUQUIM, 2006). Existe um grupo de espécies que varia bastante no gradiente de teor de areia, e que provavelmente é resultado da interação entre teor de areia e distância do curso d'água. A fertilidade do solo na área estudada esteve altamente correlacionada (de forma negativa) com teor de areia no solo, podendo essa também estar relacionada com a composição de espécies como foi descrito por outros autores (COSTA, 2004; GENTRY; EMMONS 1987; JONES et. al, 2006; VORMISTO et. al, 2004). Sendo a textura e a fertilidade do solo correlacionadas nesta área de estudo, é difícil saber qual delas de fato está afetando a composição de espécies.

A riqueza de especies em áreas proximas aos cursos d'água foi duas vezes maior que a riqueza encontrada nas áreas mais distantes (média de 25,85 e 11,72 espécies por parcelas, respectivamente). Esses resultados contrariam o estudo de Sabo et al. (2005), que reportaram para diferentes grupos biológicos, uma riqueza de espécies menor em áreas ripárias em relação às áreas circundantes. Porém o aumento no número de espécies econtrados nessas áreas corrobora com os resultados registrados por Costa et al. (2008b) para palmeiras e Drucker et al. (2008) que reportaram uma alta riqueza de espécies em áreas ripárias na Amazônia Central, bem como a ocorrência de espécies únicas de ervas nesses ambientes, sugerindo ainda que a distribuição de outros grupos taxonômicos podem apresentar fortes variações nas áreas ripárias.

Um dos mecanismos responsável pela maior riqueza de espécies de plantas é a heterogeneidade ambiental associada a áreas ripárias (GOEBEL et al., 2003; HARRIS et al., 2011; NAIMAN; DÉCAMPS, 1997), pois fornecem habitats únicos para determinados organismos (AMOROS, 2001).

Ali et al. (2000) sugerem que a umidade do solo é importante determinante da riqueza de espécies em áreas ripárias no Egito. Goebel et al. (2003) também encontraram maior riqueza de espécies herbáceas em áreas ripárias na região do Lago Norte dos Estados Unidos. A riqueza de espécies de ervas foi negativamente afetada pela falta de água em anos de secas ao longo do Rio São Pedro, voltando a aumentar quando a disponibilidade de água nessa área foi suprida através das chuvas (LITE et al., 2005). Em áreas da Amazônia meridional encontramos que baixos níveis de disponibilidade de água podem limitar a riqueza de espécies herbáceas nesses ambientes.

A luz é uma das variáveis ambientais mais relevantes para a vegetação de sub-bosque, influenciando a riqueza de espécies vegetais (ELEMANS, 2004; BARTEMUCCI et al., 2006; JELASKA et al., 2006; THOMSEN et al. 2005, BARBIER et al. 2008). O presente estudo mostrou uma relação entre as áreas ripárias, onde há maior entrada de luz, com a riqueza de espécies. Esta mesma resposta em riqueza de ervas também foi encontrada em áreas de floresta, a oeste de Chicago e do lago Michigan (BOWLES et al. 2007). Resultado semelhante para ervas e plântulas foi encontrado por Tynia et al. (2009), no oeste da Hungria, ao contrário de Hardtle et al. (2003) que observaram em seu estudo pouca influência da luminosidade sobre a riqueza de espécies de solo (plantas vasculares, briófitas e liquens). Ao contrário do presente trabalho, Zuquim et al. (2008), em seu estudo com pteridófitas na Amazônia central, observaram que a riqueza de espécies diminuiu quando houve uma abertura maior no dossel. Ervas respondem claramente a diferentes condições de disponibilidade de luz e essa heterogeneidade de micro habitats podem ser igualmente importante para riqueza de espécies (BOWLES et al., 2007, NEUFELD; YOUNG, 2003; WHIGHAM, 2004).

No sub-bosque de florestas pode haver partição do gradiente de luz, mesmo que a amplitude do gradiente seja pequena (MONTGOMERY; CHAZDON, 2002). No entanto, neste trabalho os dados de luz utilizados nas análises representam uma média por parcela, desprezando a heterogeneidade de luz ao longo dos 250 m amostrados. A detecção do efeito da luz no sub-bosque nem sempre é fácil. Além de questões técnicas, o efeito de outras variáveis ambientais sobre a riqueza de espécies estudadas também deve ser considerada (TYNIA et al., 2009).

O impacto causado pela exploração de madeira, representado pelo número de árvores cortadas em cada parcela, não afetou a composição e a riqueza de espécies como era esperado. O mesmo ocorreu em estudo sobre efeitos de exploração madeireira sobre a comunidade de ervas realizado por Costa e Magnusson (2002). O fato de que os efeitos da exploração não tenham sido significativos, não implica necessariamente que não houve nenhuma alteração na comunidade herbácea, mas que as mudanças ocorridas podem ter sido minimizadas pela regeneração, visto que muitos anos já se passaram e o dossel de florestas geralmente fecha rápido. Costa e Magnusson (2003) mostraram que a exploração seletiva teve um impacto positivo na produção de flores, frutos e no número de espécies de plantas no sub-bosque, nos primeiros anos após a exploração (5anos). Já nos 11 anos após a exploração esse impacto positivo não foi tão perceptível. Veldman et. al (2009) relataram que uma área de floresta seca na Bolivia foi afetada pela invasão de gramíneas nativas e exóticas após 5 anos de extração seletiva de madeira.

Muitas das alterações estruturais decorrentes de exploração de madeira podem ter desaparecido após longos períodos de sua ocorrência. Como a área estudada é relativamente pequena se comparada à áreas que sofrem com a exploração seletiva de madeira, que são muito maiores, não é possível generalizar a partir deste estudo que o aumento da intensidade de exploração não irá causar efeitos negativos na comunidade de ervas de sub-bosque. Entretanto, como as amostras do estudo foram bem distribuídas sobre a paisagem e cobriu um amplo gradiente de intensidade, representado pelo número de árvores cortadas (0 a 12) os resultados são robustos para afirmar que intensidades de impactos semelhantes às usadas na área de estudo não tem consequências negativas em longo prazo sobre a composição e riqueza da comunidade herbácea terrestre.

# 7. IMPLICAÇÕES PARA A CONSERVAÇÃO

A principal medida conservacionista adotada no Brasil diz respeito à criação e gestão de unidades de conservação. Os critérios utilizados para decidir que área será protegida, na maioria das vezes, não levam em conta os tipos de hábitats ou espécies protegidas.

O presente estudo buscou identificar variáveis ambientais que influencie a estrutura de comunidades de ervas de sub-bosque como um todo. O delineamento amostral desenvolvido neste trabalho, com amostragens padronizadas, foi eficiente para descrever comunidades de ervas, podendo assim contribuir para o manejo e conservação destas espécies.

Um grupo de espécies herbáceas ocorre exclusivamente em áreas ripárias e outro só ocorre em áreas mais distantes do curso d'água. Isto implica que estas áreas requerem medidas de manejo diferenciadas, para que as espécies presentes nos dois ambientes sejam conservadas. A distância do corpo d'água teve uma influência em até ~100metros na composição e riqueza de espécies no local de estudo. Estes dados tem um efeito prático quanto à conservação. Atualmente a área de preservação dos corregos prevista pelo codigo Florestal vigente é de 30m para rios de até 10m de largura. A consequência é que áreas que concentram maior riqueza de espécies na Amazônia podem estar recebendo proteção inadequada quanto aos planos de manejo. Essa área de 30m de preservação é muito estreita e algumas espécies de ervas que só ocorrem nesses ambientes serão escluídas correndo o risco de serem extintas desses locais. Nosso estudo destaca a importância dos ambientes ripários que abrigam uma mior riqueza de espécies herbáceas, sendo assim a faixa de proteção deveria ser extendida para no minímo 100 m de largura. A exploração madeireira representada pelo número de árvores cortadas não afetou a composição e a riqueza de espécies herbáceas. Logo o manejo realizado nessa área parece estar sendo realizado de maneira que mantem a comunidade herbácea, pelo menos a médio-longo prazo. O planejamento para conservação dessas áreas deve levar em conta a heterogeneidade dos gradientes ambientais interflúvio e ripário, pois as espécies não estão distribuidas de maneira homogenea, o que contemplaria a conservaçao de um maior número de espécies.

# 8. REFERÊNCIAS BIBLIOGRÁFICAS:

ANEXO 1 – Algumas das espécies encontradas nas 39 parcelas das três áreas, localizada no município de Cláudia, norte do estado de Mato Grosso.

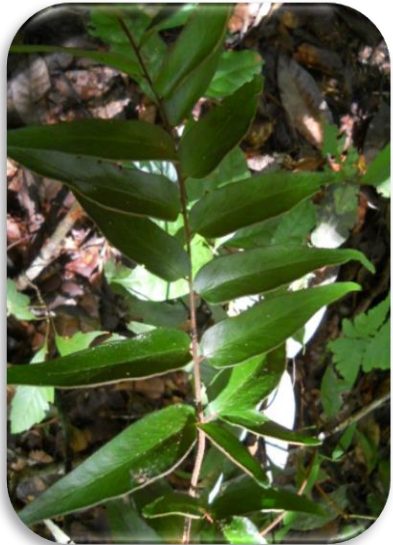
*Adiantum dolosum*

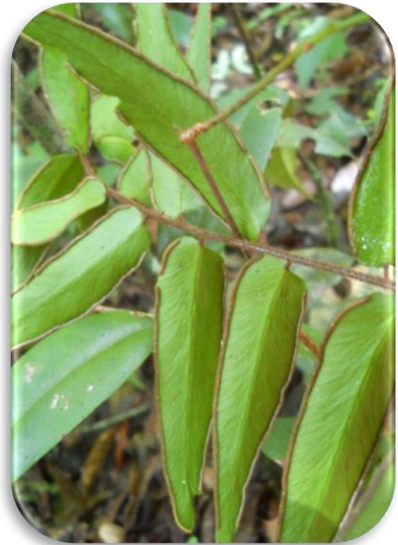
*Adiantum dolosum*

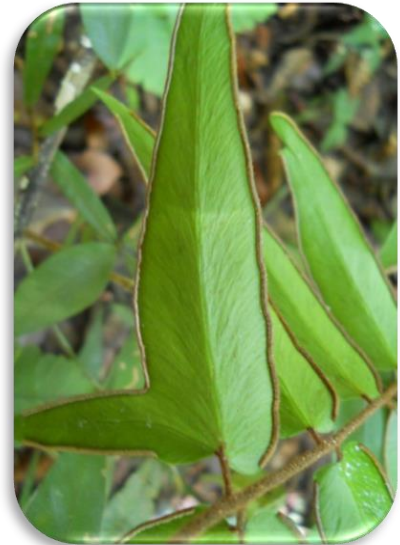
*Adiantum dolosum*

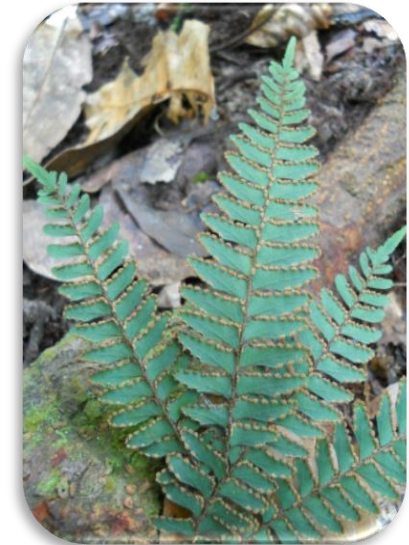
*Adiantum obliquum*

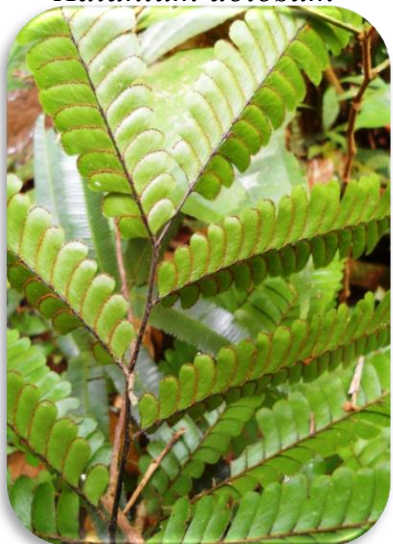
*Adiantum  tomentosum*

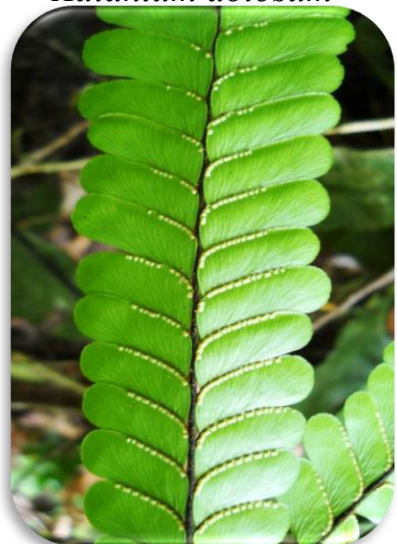
*Adiantum  tomentosum*

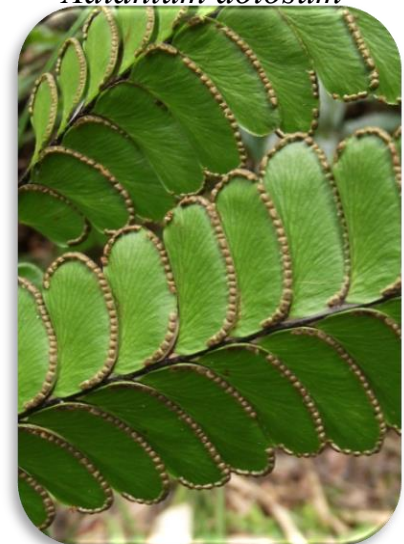
*Adiantum  tomentosum*

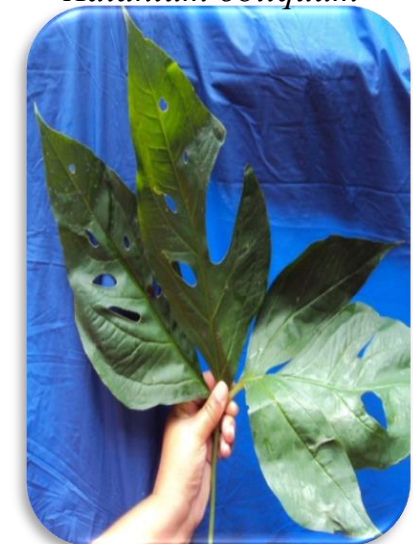
*Anaphyllopsis cururuana*

ANEXO 1 – Algumas das espécies encontradas nas 39 parcelas das três áreas, localizada no município de Cláudia, norte do estado de Mato Grosso.

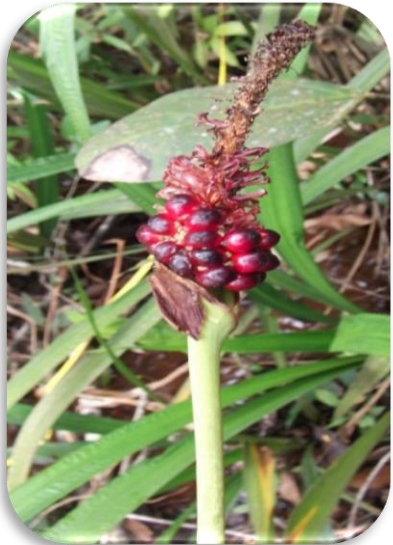
*Anaphyllopsis cururuana*

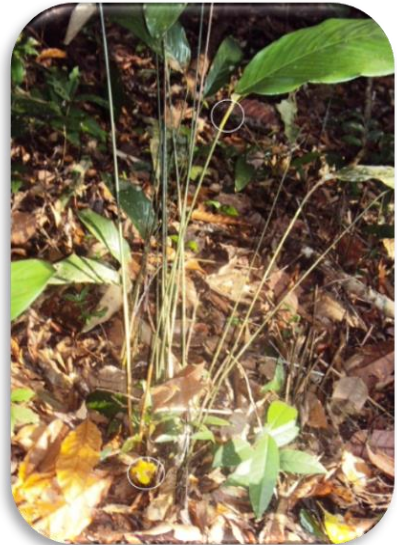
*Calathea zingiberina*

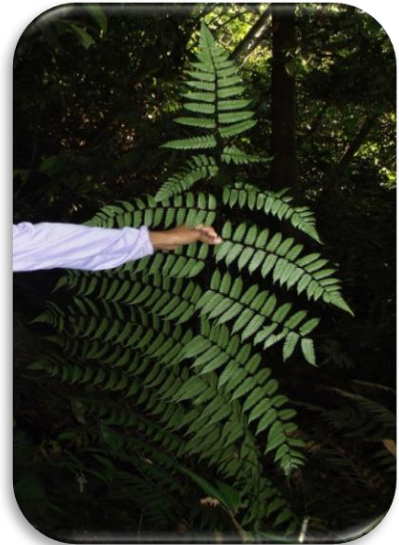
*Cyathea pungens*

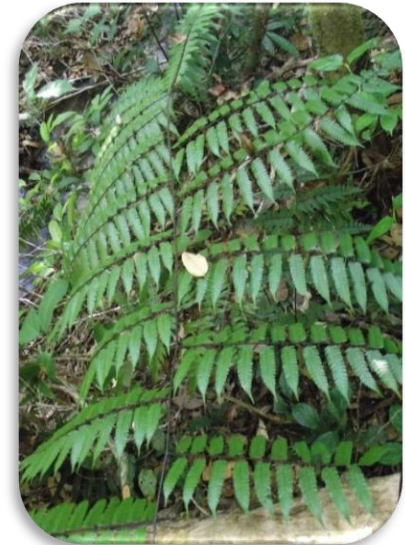
*Cyathea pungens*

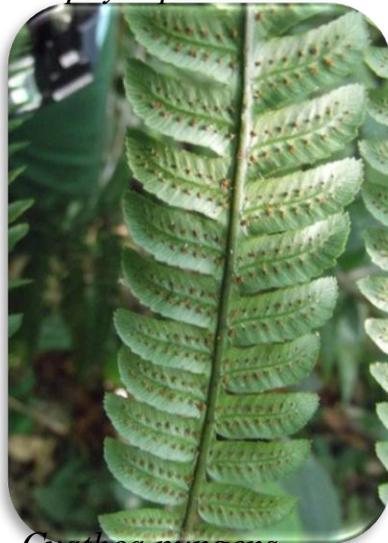
*Cyathea pungens*

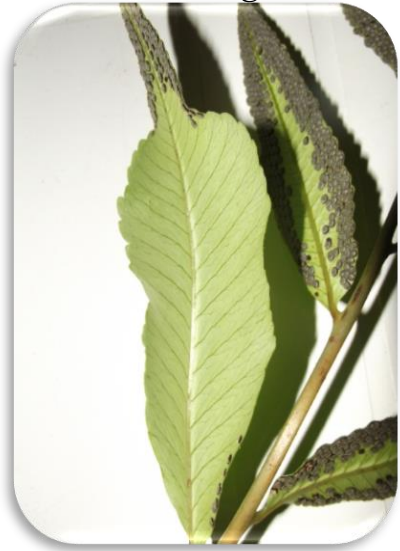
*Cyclodium meniscioides*

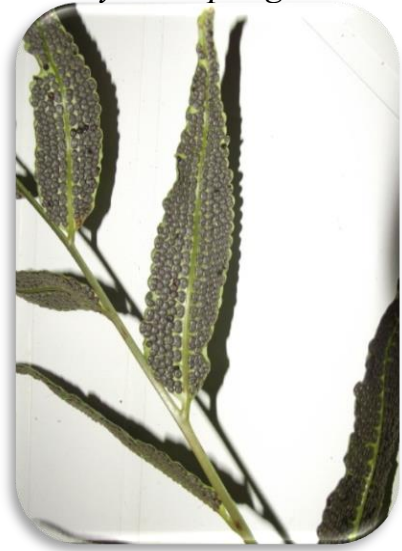
*Cyclodium meniscioides*

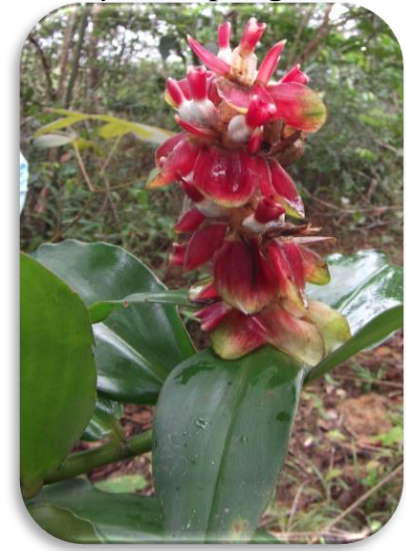
*Costus arabicus*

ANEXO 1 – Algumas das espécies encontradas nas 39 parcelas das três áreas, localizada no município de Cláudia, norte do estado de Mato Grosso

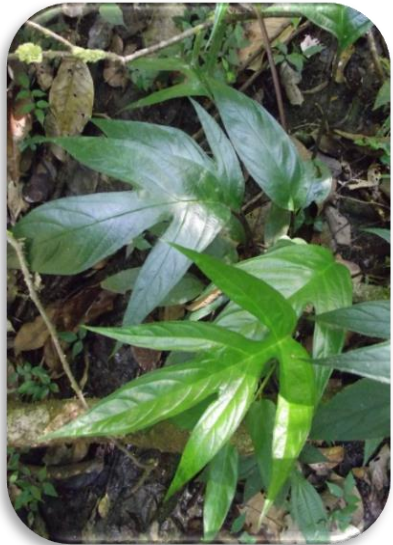
*Dorstenia* sp

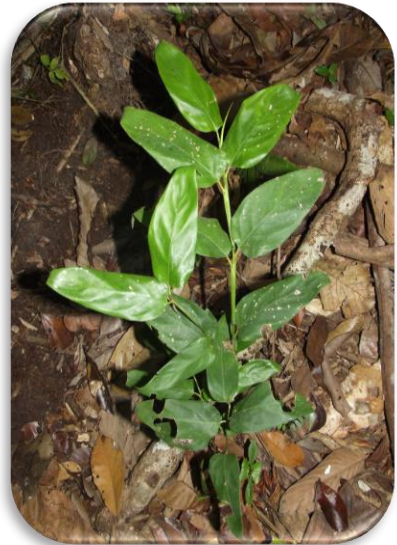
*Ischnosiphon* sp1

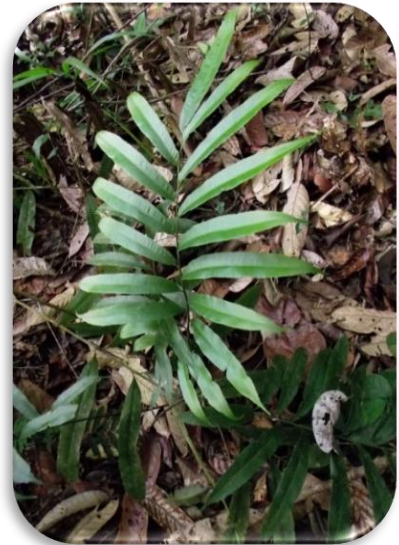
*Metaxya rostrata*

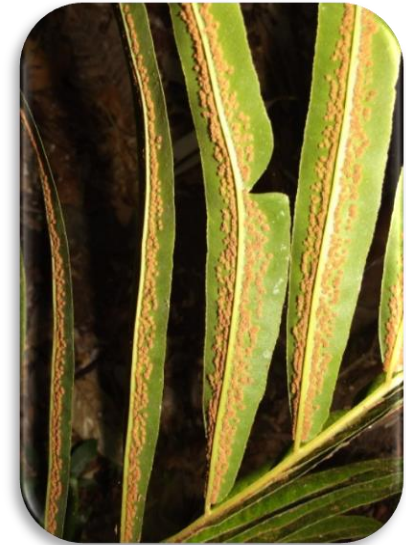
*Metaxya rostrata*

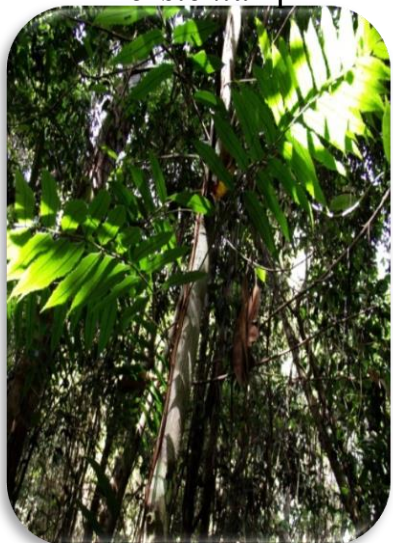
*Mickelia guianensis*

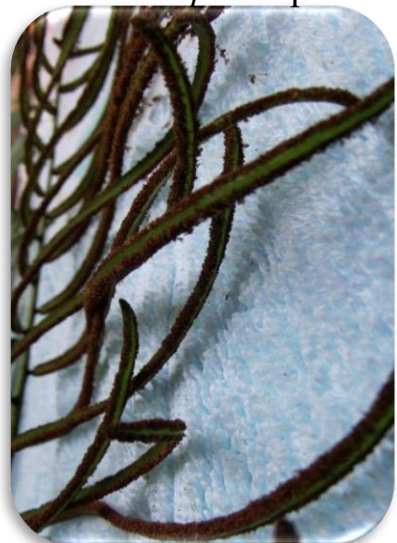
*Mickelia guianensis*

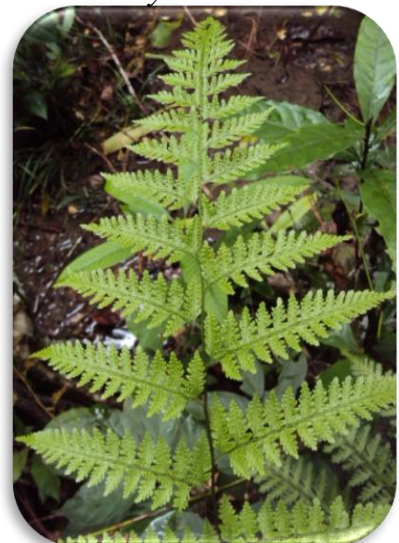
*Saccoloma inaequale*

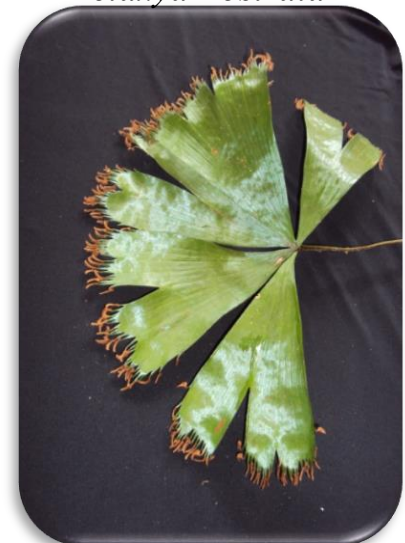
*Schizaea elegans*

ANEXO 1 – Algumas das espécies encontradas nas 39 parcelas das três áreas, localizada no município de Cláudia, norte do estado de Mato Grosso.

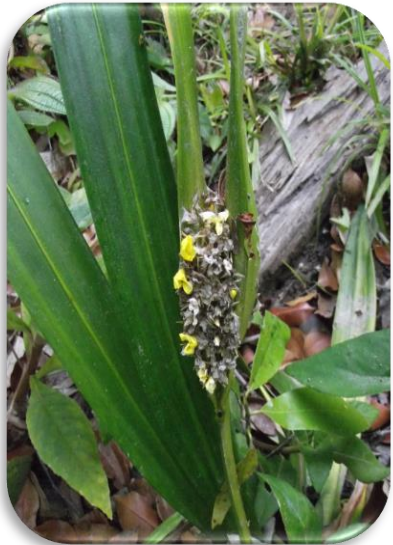
*Rapatea paludosa*

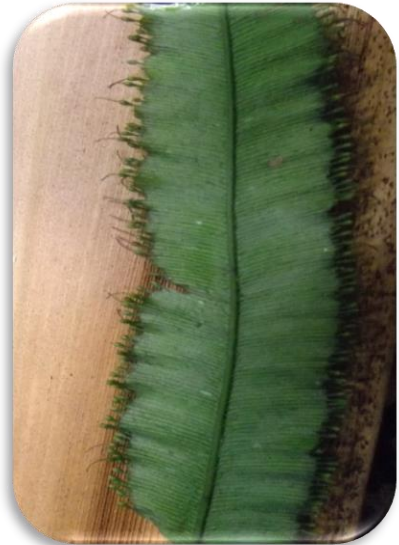
*Trichomanes vittaria*

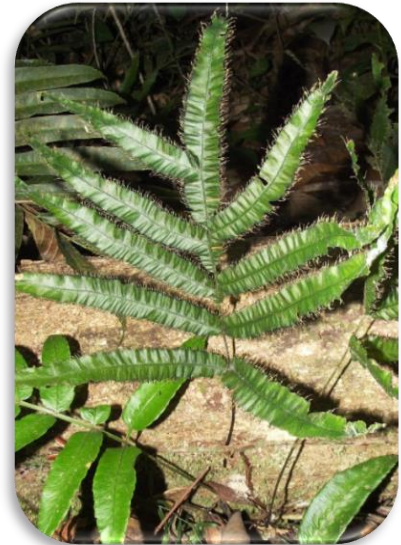
*Trichomanes pinnatum*

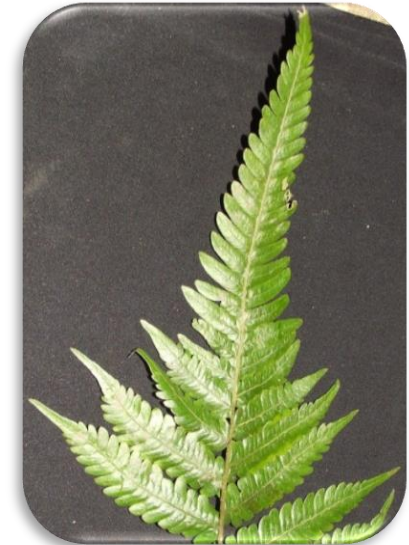
*Thelypteris leprieurii*

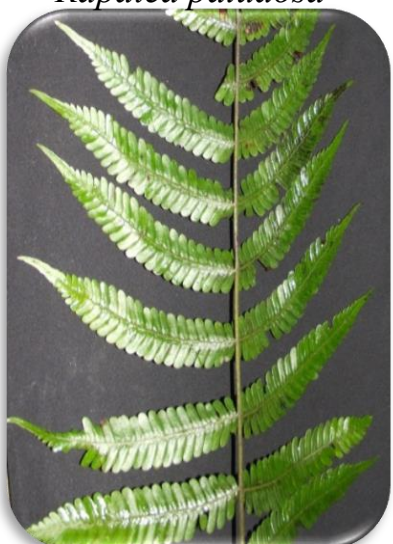
*Thelypteris leprieurii*

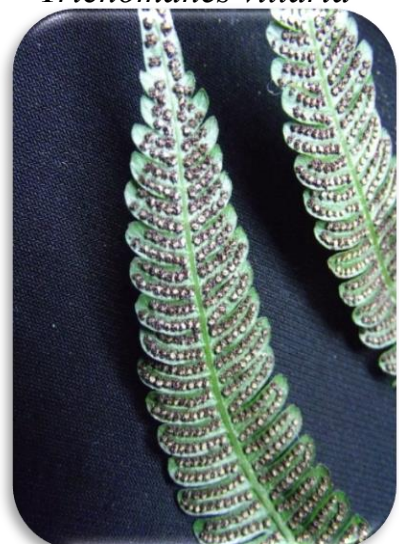
*Thelypteris leprieurii*

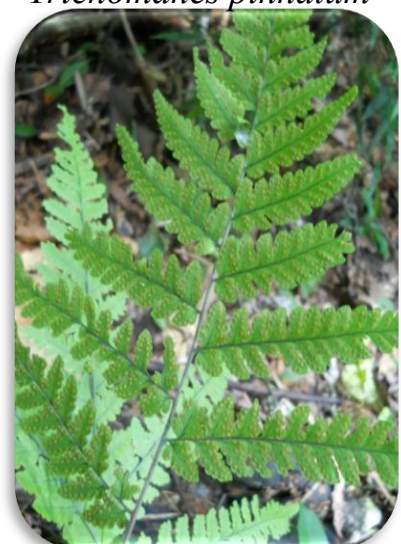
*Triplophyllum hirsutum*

Créditos: *Adiantum dolosum, Ad. obliquum, Ad. tomentosum, Mickelia guianensis, Triplophyllum hirsutum*: Miranda, R.; demais fotos: Paixão, E. C.